\documentclass[10pt,journal,letterpaper]{IEEEtran}

\usepackage{epstopdf}
\usepackage{graphicx}
\usepackage{stmaryrd}
\usepackage{amsmath,amsthm,amssymb}
\usepackage{multirow,url}
\usepackage{color,cite}
\usepackage{algorithm}
\usepackage{algorithmicx}
\usepackage{algpseudocode}
\usepackage{setspace}
\newtheorem{lem}{Lemma}
\newtheorem{thm}{Theorem}
\newtheorem{defn}{Definition}
\newtheorem{cor}{Corollary}

\newcommand{\ie}{i.e., }

\def\sgn{\mathop{\rm sgn}\nolimits}

\ifodd 0
\newcommand{\rev}[1]{{\color{blue}#1}} %revise of the text
\newcommand{\com}[1]{\textbf{\color{red} (COMMENT: #1) }} %comment of the text
\newcommand{\comg}[1]{\textbf{\color{green} (COMMENT: #1)}}
\newcommand{\response}[1]{\textbf{\color{green} (RESPONSE: #1)}} %response to comment
\else
\newcommand{\rev}[1]{#1}

\newcommand{\com}[1]{}
\newcommand{\comg}[1]{}
\newcommand{\response}[1]{}
\fi
\begin{document}

\title{Spatial Spectrum Access Game}

\author{Xu Chen, \emph{Member, IEEE}, and Jianwei Huang, \emph{Senior Member, IEEE} \thanks{Xu Chen is with the School of Electrical, Computer and Energy Engineering, Arizona State University, Tempe, Arizona, USA (email:xchen179@asu.edu). The work was mainly done when he was with the Chinese University of Hong Kong.

Jianwei Huang is with the Network Communications and Economics Lab, Department of Information Engineering, the Chinese University of Hong Kong (email:jwhuang@ie.cuhk.edu.hk). 

Part of the results have appeared in Mobihoc'12 \cite{chen2012spatial}.
}}

\maketitle

\pagestyle{empty}

\thispagestyle{empty}
\allowdisplaybreaks

%\category{C.2.1}{Network Architecture and Design}{Wireless communication}
%
%%\terms{Theory, Algorithms}
%
%\keywords{Cognitive Radio, Distributed Spectrum Sharing, Nash Equilibrium, Distributed Learning}

\begin{abstract}
A key feature of wireless communications is the spatial reuse. However, the spatial aspect is not yet well understood for the purpose of designing efficient spectrum sharing mechanisms. In this paper, we propose a framework of spatial spectrum access games on directed interference graphs, which can model quite general interference relationship with spatial reuse in wireless networks.  We show that  a pure Nash equilibrium exists for the two classes of games: (1) any spatial spectrum access games on directed acyclic graphs, and (2) any games satisfying the congestion property on directed trees and directed forests. Under mild technical conditions, the spatial spectrum access games with random backoff and Aloha channel contention mechanisms on undirected graphs also have a pure Nash equilibrium. We also quantify the price of anarchy of the spatial spectrum access game. We
then propose a distributed learning algorithm, which only utilizes users' local observations to adaptively adjust the spectrum access strategies.  We show that the distributed learning algorithm can converge to an approximate mixed-strategy Nash equilibrium for any spatial spectrum access games.  Numerical results demonstrate that the distributed learning algorithm achieves up to $100\%$ performance improvement over a random access algorithm.
\end{abstract}
\section{Introduction}
Cognitive radio is envisioned as a promising technology to alleviate
the problem of spectrum under-utilization \cite{key-2}. It enables unlicensed
wireless users (secondary users) to opportunistically access the licensed
channels owned by legacy spectrum holders (primary users), and thus
can significantly improve the spectrum efficiency \cite{key-2}.

A key challenge of the cognitive radio technology is how to resolve the resource competition
by selfish secondary users in a decentralized fashion. If multiple
secondary users transmit over the same channel simultaneously, severe interferences or collisions might occur and the individual as well as total data rates of all users may get reduced. Therefore, it is necessary
to design efficient spectrum sharing mechanisms for cognitive radio
networks.

\begin{figure}[tt]
\centering
\includegraphics[scale=0.45]{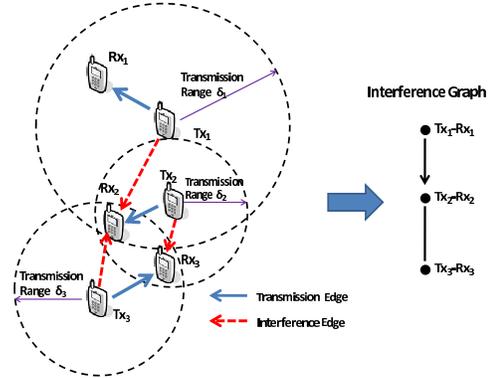}
\caption{\label{fig:Distributed-spectrum-access}Illustration of distributed
spectrum access with spatial reuse under the protocol interference model. Each user $n$ is represented by a transmitter $Tx_{n}$ and receiver $Rx_{n}$ pair. Users $2$ and $3$ can not generate interference to user $1$, since user $1$'s receiver $Rx_{1}$ is far from user $2$ and $3$'s transmitters. On the other hand,  user $1$ can generate interference to user $2$, since user $2$'s receiver $Rx_{2}$ is within the transmission range of user $1$'s transmitter $Tx_{1}$. Similarly, user $2$ and user $3$ can generate interferences to each other.}
\end{figure}

The competitions among secondary users for common spectrum have often
been studied as a noncooperative game (e.g., \cite{key-3,key-21,key-4,key-22}  and the references therein). For example, Nie and Comaniciu in \cite{key-21} designed a self-enforcing distributed
spectrum access mechanism based on potential games. Niyato and Hossain
in \cite{key-4} studied a price-based spectrum access mechanism for
competitive secondary users. F¨¦legyh¨¢zi \emph{et al.} in \cite{key-22} proposed a
two-tier game framework for medium access control (MAC) mechanism
design.

When not knowing the spectrum information such as channel availability, secondary users need to learn the network environment and adapt the spectrum access decisions accordingly. Han \emph{et al.} in \cite{key-25} used the no-regret learning to solve this problem, assuming that the users' channel selections are common information. When users' channel selections are not observable, Anandkumar \emph{et al.}  in \cite{key-27} as well as Liu and Zhao in \cite{key-29} designed multi-agent multi-armed bandit learning algorithms to minimize the expected performance loss of distributed spectrum access.

%\vspace{-0.05in}

A common assumption of the above studies  is that secondary users are
close-by and interfere with each other when they transmit on the same
channel simultaneously. However, a unique feature of wireless communication is spatial reuse. If users who transmit simultaneously
are located sufficiently far away, then simultaneous transmissions over the same channel may not cause any
performance degradation to the users. Such spatial effect on spectrum sharing
is less understood than many other aspects in the existing literature \cite{key-60}.

\begin{figure}[tt]
\centering
\includegraphics[scale=0.42]{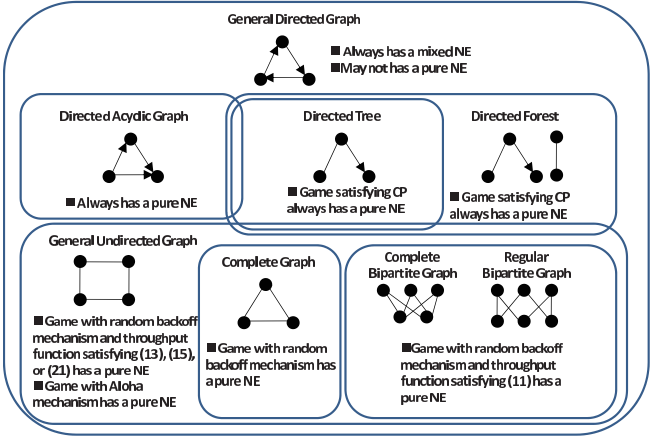}
\caption{\label{fig:Summary}Summary of the results on the existence of both mixed and pure Nash equilibrium (NE) of spatial spectrum access games. Here CP means congestion property (see Section \ref{NE}). }
\end{figure}

%The rest of the paper is organized as follows. We introduce the system
%model and the spatial spectrum  access game in Sections \ref{sec:System-Model} and \ref{game}, respectively. We investigate the existence of Nash equilibria in Section \ref{NE}. Then we present the distributed learning algorithm in Section \ref{Learning}.
%We illustrate the performance of the proposed algorithm through numerical
%results in Section \ref{sec:Numerical-Results}, and finally conclude
%in Section \ref{sec:Conclusion}.

Recently, Tekin \emph{et al.} in \cite{tekinatomic} proposed a novel spatial
congestion game framework to take spatial relationship into account.
The key idea is to extend the classical congestion game upon an \emph{undirected} graph, by assuming that the interferences among the players are symmetric and  a player's throughput depends on the number of players in its neighborhood that choose the same resource. However, the interference relationship among the secondary users can be asymmetric due to the heterogeneous transmission powers and locations of the users. We hence propose a more general framework of spatial spectrum access game on \emph{directed} interference graphs, which take users' heterogeneous resource competition capabilities and asymmetric interference relationship into account. The congestion game on directed graphs has also been studied in \cite{bil¨°2008graphical}, with the assumption that players have linear and homogeneous payoff functions. The game model in this paper is more general  and allows both linear and nonlinear player-specific payoff functions. Moreover, we design a distributed algorithm for achieving the equilibria of the game. The main results and
contributions of this paper are as follows:
\begin{itemize}
\item \emph{General game formulation}: We formulate the distributed spectrum access problem as a spatial spectrum access game on directed interference graphs, with user-specific channel data rates and channel contention capabilities.
\item \emph{Existence of Nash equilibria}: We show by counter examples  that a general spatial spectrum access game may not have a pure Nash equilibrium. We then show that  a pure strategy equilibrium exists in the following two classes of games: (1) any spatial spectrum access games on directed acyclic graphs, and (2) any games satisfying the congestion property on directed trees and directed forests. We also show that under mild conditions the spatial spectrum access games with random backoff and Aloha channel contention mechanisms on undirected graphs are potential games and have pure Nash equilibria.  We also quantify the price of anarchy of the spatial spectrum access game.
\item \emph{Distributed learning for achieving an approximate Nash equilibrium}: We develop a maximum likelihood estimation approach for estimating user expected throughput based on local observations. Based on  the local estimation of user expected throughputs, we then propose a distributed learning algorithm that can converge to an approximate mixed Nash equilibrium for any spatial spectrum access games.  Numerical results demonstrate that the distributed learning algorithm achieves up-to $100\%$ performance improvement over the random access algorithm.
\end{itemize}

Due the space limit, all the proofs in this paper can be found in the supplementary file on the TMC website or the online technical report in \cite{SSAG}.

\section{System Model}\label{sec:System-Model}

We consider a cognitive radio network with a set  $\mathcal{M}=\{1,2,...,M\}$ of independent and stochastically
heterogeneous primary channels. A set $\mathcal{N}=\{1,2,...,N\}$ of secondary users try to access these channels distributively when the channels are not occupied by primary (licensed) transmissions. Here we assume that each secondary user is a dedicated transmitter-receiver pair.

To take users' spatial relationship into account, we denote $\boldsymbol{d}_{n}=(d_{Tx_{n}},d_{Rx_{n}})$ as the \textbf{location vector} of secondary user $n$, where $d_{Tx_{n}}$ and $d_{Rx_{n}}$ denote the location of the transmitter and the receiver, respectively. Each secondary user $n$ has a \textbf{interference range} $\nu_{n}$. Then
given the location vectors of all secondary users, we can obtain the \textbf{interference
graph} $G=\{\mathcal{N},\mathcal{E}\}$ to describe the interference relationship among the users. Here the vertex set $\mathcal{N}$
is the same as the secondary user set. The edge set  is defined as $\mathcal{E}=\{(i,j):||d_{Tx_{i}},d_{Rx_{j}}||\leq\nu_{i},\forall i,j\neq i\in\mathcal{N}\}$, where  $||d_{Tx_{i}},d_{Rx_{j}}||$ is the
distance between the transmitter of user $i$ and  the receiver of user $j$. In general, an interference edge can be directed or undirected. If an interference edge is directed from secondary user $i$ to user $j$, then user $j$'s data transmission will be affected by user $i$'s transmission on the same channel, but user $i$ will not be affected by user $j$. If the interference edge is undirected\footnote{Here the edge is actually bi-directed. We follow the conventions in \cite{tekinatomic} and ignore the directions on the edge.}  between user $i$ and user $j$, then the two users can affect each other. Note that a generic directed interference graph can consist of a mixture of directed and undirected edges. In the sequel, we call an interference graph undirected, if and only if all the edges of the graph are undirected.  We also denote the set of users that can cause interference to user $n$ as $\mathcal{N}_{n}=\{i:(i,n)\in\mathcal{E},i\in\mathcal{N}\}$.

Based on the interference model above, we describe the cognitive radio network with a slotted transmission structure as follows:
\begin{itemize}
\item \emph{Channel State}: the channel state for a channel $m$ during
time slot $t$ is $S_{m}(t)=0$ if channel \ensuremath{m} is occupied by primary transmissions and $S_{m}(t)=1$ otherwise.

\item \emph{Channel State Transition}: for a channel $m$, the channel state $S_{m}(t)$ is a random variable with a probability density
function as $\psi_{m}$. In the following, we denote the channel idle probability $\theta_{m}$ as the mean of $S_{m}(t)$, i.e., $\theta_{m}=E_{\psi_{m}}[S_{m}(t)]$. For example, the state of a channel changes according to a two-state Markovian process. We denote the channel state probability
vector of channel $m$ at time $t$ as $
\boldsymbol{q}_{m}(t)\triangleq(Pr\{S_{m}(t)=0\},Pr\{S_{m}(t)=1\}),$
 which forms a Markov chain as $
\boldsymbol{q}_{m}(t)=\boldsymbol{q}_{m}(t-1)\Gamma_{m},\forall t\geq 1,$
 with the transition matrix \[
\Gamma_{m}=\left[\begin{array}{cc}
1-\varepsilon_{m} & \varepsilon_{m}\\
\xi_{m} & 1-\xi_{m}\end{array}\right].\]
Furthermore, the long run statistical channel availability $\theta_{m}\in(0,1)$ of a channel $m$ can be obtained from the stationary distribution
of the Markov chain, i.e., \begin{align}
\theta_{m}=\frac{\varepsilon_{m}}{\varepsilon_{m}+\xi_{m}}.\label{eq:sd-2}\end{align}

As another example, we can also use the channel idle probability $\theta_{m}$ to indicate the spectrum availability for white-space spectrum access. Due to the fact that the activities of primary users on TV channels typically change very slowly, the most recent FCC ruling requires white-space devices (i.e., secondary users of TV channels) to determine the spectrum availability via a database \cite{FCC}. In this case, we can set that $\theta_{m}=1$ if the TV channel $m$ is vacant for secondary users and $\theta_{m}=0$ otherwise.

%\begin{figure}[tt]
%\begin{center}
%\includegraphics[scale=0.5]{MarkovChannel}
%\caption{\label{fig:Markovian-Channel-Model}Two states Markovian channel model}
%\end{center}
%\end{figure}

\item \emph{User Specific Channel Throughput}: for each secondary
user $n$, its realized data rate $b_{m}^{n}(t)$
on an idle channel $m$ in each time slot evolves according
to a random process with a mean $B_{m}^{n}$, due to users'
heterogeneous transmission technologies and the local environmental
effects such as fading. For example, we can compute the data rate
$b_{m}^{n}(t)$ according to the Shannon capacity as\begin{equation}
b_{m}^{n}(t)=W\log_{2}\left(1+\frac{\eta_{n}z_{m}^{n}(t)}{\omega_{m}^{n}}\right),\label{eq:dd}\end{equation}
where $W$ is the channel bandwidth, $\eta_{n}$ is
the fixed transmission power adopted by user $n$ according to the requirements such as the primary user protection, $\omega_{m}^{n}$ denotes
the background noise power, and $z_{m}^{n}(t)$ is the channel gain. In a Rayleigh fading
channel environment, the channel gain $z_{m}^{n}(t)$ is a realization
of a random variable that follows the exponential distribution.
\item \emph{Time Slot Structure}: each secondary user $n$ executes the
following stages synchronously during each time slot:
\begin{itemize}
\item \emph{Channel Sensing}: sense one of the channels based on the channel
selection decision made at the end of previous time slot.
\item \emph{Channel Contention}: Let
$a_{n}$ be the channel selected by user $n$, and $\boldsymbol{a}=(a_{1},...,a_{N})$ be the channel selection profile of all users. The probability
that user $n$ can grab the chosen idle channel $a_{n}$ during a
time slot is $g_{n}(\mathcal{N}_{n}^{a_{n}}(\boldsymbol{a}))\in(0,1)$, which depends
on the subset of user $n$'s interfering users that choose the same channel $\mathcal{N}_{n}^{a_{n}}(\boldsymbol{a})\triangleq\{i\in\mathcal{N}_{n}:a_{i}=a_{n}\}$. Here are two examples: \\
%\begin{itemize}
1) \emph{Random backoff mechanism}: the contention stage of a time slot is divided
into $\lambda_{\max}$ mini-slots\footnote{The contention window size $\lambda_{\max}$ plays an important role for optimizing the system performance. If $\lambda_{\max}$ is too small, it would increase the collision probability among users and hence negatively affect the system performance. In general, if $\lambda_{\max}$ is too large, it would reduce the spectrum access time and hence reduce the system throughput.  To optimize the system performance, we can adopt the approach in \cite{anouar2007optimal} to determine the optimal contention window size.}. Each contending user $n$ first
counts down according to a randomly and uniformly generated integer backoff
time counter (number of mini-slots) $\lambda_{n}$ between $1$ and $\lambda_{\max}$.
If there is no active transmissions till the count-down timer expires, the user monitors the channel and transmits RTS/CTS messages on that channel.
If multiple users choose the same backoff counter, a collision will
occur and no users can grab the channel successfully.
Once successfully gets the channel, the user starts to transmit
its data packet. In this case, we have\begin{align}
   & g_{n}(\mathcal{N}_{n}^{a_{n}}(\boldsymbol{a}))=Pr\{\lambda_{n}<\min_{i\in\mathcal{N}_{n}:a_{i}=a_{n}}\{\lambda_{i}\}\}\nonumber \\
 = & \sum_{\lambda=1}^{\lambda_{\max}}Pr\{\lambda_{n}=\lambda\}Pr\{\lambda_{n}<\min_{i\in\mathcal{N}_{n}:a_{i}=a_{n}}\{\lambda_{i}\}|\lambda_{n}=\lambda\} \nonumber \\
  = & \sum_{\lambda=1}^{\lambda_{\max}}\frac{1}{\lambda_{\max}}\left(\frac{\lambda_{\max}-\lambda}{\lambda_{\max}}\right)^{K_{n}^{a_{n}}(\boldsymbol{a})},\label{eq:t1}\end{align}
where $K_{n}^{a_{n}}(\boldsymbol{a})=|\mathcal{N}_{n}^{a_{n}}(\boldsymbol{a})|=\sum_{i\in\mathcal{N}_{n}}I_{\{a_{i}=a_{n}\}}$
denotes the number of user $n$'s interfering users choosing the same channel
as user $n$. \\
2) \emph{Aloha mechanism}: user $n$ contends for an idle channel with a probability
$p_{n}\in(0,1)$ in a time slot. If multiple interfering users
contend for the same channel, a collision occurs and no user can grab
the channel for data transmission. In this case, we have\begin{eqnarray}
 g_{n}(\mathcal{N}_{n}^{a_{n}}(\boldsymbol{a}))& = & p_{n}\prod_{i\in\mathcal{N}_{n}^{a_{n}}(\boldsymbol{a})}\left(1-p_{i}\right).\label{eq:t2}\end{eqnarray}
%\end{itemize}
%Note that for the random backoff mechanism, the channel grabbing probability
%$g_{n}(\mathcal{N}_{n}^{a_{n}}(\boldsymbol{a}))$ is \emph{user homogeneous}
%since it only depends on the number of contending users $K_{n}^{a_{n}}(\boldsymbol{a})$.
%For the Aloha mechanism, the channel grabbing probability $g_{n}(\mathcal{N}_{n}^{a_{n}}(\boldsymbol{a}))$
%is \emph{user heterogeneous} since it depends on who (instead of how many users) contend
%the channel.
\item \emph{Data Transmission}: transmit data packets if the user successfully
grabs the channel.
\item \emph{Channel Selection}: choose a channel to access during next time slot
according to the distributed learning algorithm in Section \ref{Learning}.
\end{itemize}
\end{itemize}

%\begin{figure}[tt]
%\centering
%\includegraphics[scale=0.5]{time_slot}
%\caption{\label{fig:Time-slot-structure}Time slot structure with random backoff mechanism}
%\end{figure}

Under a fixed channel selection profile $\boldsymbol{a}$, the long-run average throughput of a secondary user
$n$ choosing channel $a_{n}$ can be computed as\begin{equation}
U_{n}(\boldsymbol{a})=\theta_{a_{n}}B_{a_{n}}^{n}g_{n}(\mathcal{N}_{n}^{a_{n}}(\boldsymbol{a})).\label{eq:u1}\end{equation}

\rev{Note that in practices, due to hardware constraint each secondary user typically cannot observe the channel states of all the channels in each time slot. In this case, one possible modeling approach is to formulate the distributed spectrum access problem as a partial-observation dynamic game, such that the game state is defined as the channel states $\{S_m(t)\}_{m=1}^{M}$ in the current time slot, and each secondary user has a partial observation of the game state in each time slot. However, it is well-known that such a partial-observation dynamic game is very difficult to analyze, and is computationally intractable due to the curse of dimensionality. To enable tractable analysis and achieve an efficient spectrum access, we hence study the distributed spectrum access problem from the long-run average perspective, and utilize the statistical channel availability information (i.e., channel idle probability $\theta_m$) to aid secondary users' decisions makings. This is because that the statistical channel availability information can be learned from the history of a secondary user's local observations, via the maximum likelihood estimation approach described in Section \ref{Learning}. Moreover, the spatial spectrum access game solution in this paper can help us solve  the complete information dynamic game, where each secondary user is able to globally observe  the channel state realization $\{S_m(t)\}_{m=1}^{M}$ of all the channels in each time slot. Since the secondary users cannot control the transition of channel states, we can easily derive the solution of the complete information dynamic game as follow: for each time slot $t$, we solve the corresponding stage spatial spectrum access game using the algorithm proposed in this paper,  with the channel idle probabilities $\{\theta_m\}_{m=1}^{M}$ of the stage game replaced by  the channel state realization $\{S_m(t)\}_{m=1}^{M}$ of time slot $t$.} Since our analysis is from the secondary users\textquoteright{} perspective,
we will use the terms \textquotedblleft{}secondary user\textquotedblright{}
and \textquotedblleft{}user\textquotedblright{} interchangeably.

\section{Spatial Spectrum Access Game}\label{game}

We now consider the problem that each user tries to maximize its own
throughput by choosing a proper channel distributively. Let $a_{-n}=\{a_{1},...,a_{n-1},a_{n+1},...,a_{N}\}$
be the channels chosen by all other users except user $n$. Given
other users' channel selections $a_{-n}$, the problem faced by a user
$n$ is\begin{equation}
\max_{a_{n}\in\mathcal{M}}U(a_{n},a_{-n}),\forall n\in\mathcal{N}.\label{eq:s1}\end{equation}
The distributed nature of the channel selection problem naturally
leads to a formulation based on the game theory, such that users
can self organize into a mutually acceptable channel selection (\textbf{pure
Nash equilibrium}) $\boldsymbol{a}^{*}=(a_{1}^{*},a_{2}^{*},...,a_{N}^{*})$
with\begin{equation}
a_{n}^{*}=\arg\max_{a_{n}\in\mathcal{M}}U(a_{n},a_{-n}^{*}),\forall n\in\mathcal{N}.\label{eq:s2}\end{equation}
We thus formulate the distributed channel selection problem on an
interference graph $G$ as a \textbf{spatial spectrum access game} $\Gamma=(\mathcal{N},\mathcal{M},G,\{U_{n}\}_{n\in\mathcal{N}})$,
where $\mathcal{N}$ is the set of players, $\mathcal{M}$ is the set
of strategies, $G$ describes the interference relationship among
the players, and $U_{n}$ is the payoff function of player $n$.

It is known that not every finite strategic game possesses a pure Nash equilibrium \cite{Nash}. We then introduce a more general concept of mixed Nash equilibrium. Let $\boldsymbol{\sigma}_{n}\triangleq(\sigma_{1}^{n},...,\sigma_{M}^{n})$
denote the mixed strategy of user $n$, where $0\leq\sigma_{m}^{n}\leq1$
is the probability of user $n$ choosing channel $m$, and $\sum_{m=1}^{M}\sigma_{m}^{n}=1$. For simplicity, we use the
same payoff notation $U_{n}(\boldsymbol{\sigma}_{1},...,\boldsymbol{\sigma}_{N})$ to
denote the expected throughput of user $n$ under the mixed strategy
profile $(\boldsymbol{\sigma}_{1},...,\boldsymbol{\sigma}_{N})$,
and it can be computed as\begin{equation}
U_{n}(\boldsymbol{\sigma}_{1},...,\boldsymbol{\sigma}_{N})=\sum_{a_{1}=1}^{M}\sigma_{a_{1}}^{1}...\sum_{a_{N}=1}^{M}\sigma_{a_{N}}^{N}U_{n}(a_{1},...,a_{N}).\label{eq:s3}\end{equation}
Similarly to the pure Nash equilibrium, the mixed Nash equilibrium is
defined as:
\begin{defn}[\textbf{Mixed Nash Equilibrium} \!\!\cite{Nash}]
The mixed strategy profile $\boldsymbol{\sigma}^{*}=(\boldsymbol{\sigma}_{1}^{*},...,\boldsymbol{\sigma}_{N}^{*})$
is a mixed Nash equilibrium, if for every user $n\in\mathcal{N}$,
we have\[
U_{n}(\boldsymbol{\sigma}_{n}^{*},\boldsymbol{\sigma}_{-n}^{*})\geq U_{n}(\boldsymbol{\sigma}_{n},\boldsymbol{\sigma}_{-n}^{*}),\forall\boldsymbol{\sigma}_{n}\neq\boldsymbol{\sigma}_{n}^{*},\]
where $\boldsymbol{\sigma}_{-n}^{*}$ denote the mixed strategy choices
of all other users except user $n$.
\end{defn}

One important property deduced from the mixed Nash equilibrium definition is that  if a user assigns some positive probabilities of choosing some actions, then the expected payoff of these actions should be the same at the equilibrium. Otherwise, the user can improve by increasing the probability of choosing the action with higher expected payoff. In the spatial spectrum access game, each secondary user takes both the primary activity levels on different channels and the competition with other secondary users into consideration, in order to improve its long run average throughput.  Nash equilibrium is the natural solution concept  for the spatial spectrum access game. At a Nash equilibrium, secondary users are mutually satisfied with their long run average throughputs and no user can improve by changing its channel unilaterally.

In the following sections, we will study the existence of both mixed and pure Nash equilibria based on the global network information, and then discuss how to achieve the Nash equilibria based on the local user observations only.

\section{Existence of Nash Equilibria}\label{NE}

In this part, we study the existence of Nash equilibria in a spatial spectrum access game. Since a spatial spectrum access game is a finite strategic game  (i.e., with finite number of players and finite number of channels), we know that it always admits a mixed Nash equilibrium according to \cite{Nash}.

%\begin{thm}\label{MixNE}
%Every spatial spectrum access game on a directed interference
%graph has a mixed Nash equilibrium.\end{thm}

On the other hand, not every finite strategic game possesses a pure Nash equilibrium \cite{Nash}. Compared with mixed Nash equilibrium, Pure Nash equilibrium can achieve the mutually satisfactory spectrum sharing solution without requiring the frequent channel switching and hence helps to reduce the system overhead such as energy consumption of frequent channel switching. This motivates us to further investigate the existence of pure Nash Equilibria of the spatial spectrum access games.

\subsection{Existence of Pure Nash Equilibria on Directed Interference Graphs}\label{directed}
We first study the existence of pure Nash Equilibria on directed interference graphs.

First of all, we can construct a game which  does not have a pure Nash equilibrium.

\begin{thm}
There exists a spatial spectrum access game on a directed interference
graph not admitting any pure Nash equilibrium.
\end{thm}

\begin{figure}[tt]
\centering
\includegraphics[scale=0.45]{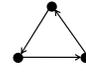}
\caption{\label{fig:An-example-of}An example of spatial spectrum access game
without pure Nash equilibria. There are two channels available and
the throughput of a user $n$ is given as $U_{n}(\boldsymbol{a})=p\prod_{i\in\mathcal{N}_{n}^{a_{n}}(\boldsymbol{a})}(1-p)$. If all three players (nodes) choose channel $1$, then each player has the incentive of choosing channel $2$ to improve its throughput assuming that the other two players do not change their channel choices. We can show that such derivation will happen for all $8$ possible strategy profiles $\boldsymbol{a}=(a_1,a_2,a_3)$, where $a_i\in\{1,2\}$ for $i\in\{1,2,3\}$.}
\end{figure}

Figure \ref{fig:An-example-of} shows such an example. It is easy to verify that
for all $8$ possible channel selection profiles, there always exists
one user (out of these three users) having an incentive to change
its channel selection unilaterally  to improve its throughput.

%\begin{figure}[tt]
%\begin{minipage}[t]{0.48\linewidth}
%\centering
%\includegraphics[scale=0.7]{3Graph}
%\caption{\label{fig:An-example-of}An example of spatial spectrum access game
%without pure Nash equilibria. There are two channels available and
%the throughput of a user $n$ is given as $U_{n}(\boldsymbol{a})=p\prod_{i\in\mathcal{N}_{n}^{a_{n}}(\boldsymbol{a})}(1-p)$. If all three players (nodes) choose channel $1$, then each player has the incentive of choosing channel $2$ to improve its throughput assuming that the other two players do not change their channel choices. We can show that such derivation will happen for all $8$ possible strategy profiles $\boldsymbol{a}=(a_1,a_2,a_3)$, where $a_i\in\{1,2\}$ for $i\in\{1,2,3\}$.}
%\end{minipage}
%\hfill
%\begin{minipage}[t]{0.48\linewidth}
%\centering
%\includegraphics[scale=0.45]{Hgraph}
%\caption{\label{fig:A-graph-consists}An interference graph that consists of directed acyclic
%graphs and directed trees}
%\end{minipage}
%\end{figure}

We then focus on identifying the conditions under which the game admits a pure Nash equilibrium. To proceed, we first introduce the following lemma (the proof is given in the appendix in the separate supplemental file).
\begin{lem}
\label{lem:If-any-spatial-1}Assume that any spatial spectrum access game with $N$ users on
a given directed interference graph $G$ has a pure Nash equilibrium. Then
we can construct a new spatial spectrum access game by adding a new player, who
can not generate interference to any player in the original game and
may receive interference from one or multiple players in the original
game. The new game with $N+1$ users also has a pure Nash equilibrium.\end{lem}
%\begin{proof}
%Suppose that the original spatial spectrum access game on the interference
%graph $G$ has a pure Nash equilibrium $\boldsymbol{a}^{*}.$ We index the newly
%added player as player $N+1$, denote the set of players that can
%generate interference to player $N+1$ as $\mathcal{I}(\boldsymbol{a}^{*})$, and denote
%the set of player $N+1$'s interfering players that choose
%channel $m$ as $\mathcal{I}_{m}(\boldsymbol{a}^{*})\triangleq\{n\in\mathcal{I}(\boldsymbol{a}^{*}):a_{n}^{*}=m\}$.
%Now player $N+1$ can compute its best response strategy $a_{N+1}^{*}=\arg\max_{a\in\mathcal{M}}\{\theta_{a}B_{a}^{N+1}g_{N+1}(\mathcal{I}_{a}(\boldsymbol{a}^{*}))\}$.
%Obviously, the strategy profile $(\boldsymbol{a}^{*},a_{N+1}^{*})$ of $N+1$ players
%in the new game is a pure Nash equilibrium.
%\end{proof}

We know that any directed acyclic graph (i.e., a directed
graph contains no directed cycles) can be given a topological sort (i.e., an ordering of the nodes), such that if node $i<j$ then there
are no edges directed from the node $j$ to node $i$ in the ordering \cite{Graph}.
This is due to that any spatial spectrum access game with $N=1$ user always has a pure Nash equilibrium. By the induction argument,   we know from Lemma \ref{lem:If-any-spatial-1} that
\begin{cor}
Any spatial spectrum access game on a directed acyclic graph has a
pure Nash equilibrium.
\end{cor}

To obtain more insightful results, we next impose the following property on the spatial spectrum access games:
\begin{defn}[\textbf{Congestion Property}] \label{def:cp}
User $n$'s channel grabbing probability $g_{n}\left(\mathcal{N}_{n}^{a_{n}}(\boldsymbol{a})\right)$ satisfies the congestion property if for any $\tilde{\mathcal{N}}_{n}^{a_{n}}(\boldsymbol{a})\subseteq\mathcal{N}_{n}^{a_{n}}(\boldsymbol{a})$, we have
\begin{align}g_{n}(\tilde{\mathcal{N}}_{n}^{a_{n}}(\boldsymbol{a}))\geq g_{n}\left(\mathcal{N}_{n}^{a_{n}}(\boldsymbol{a})\right).\label{CP}\end{align}
Furthermore, a spatial spectrum access game satisfies the congestion property if (\ref{CP}) holds for all users $n\in\mathcal{N}.$
\end{defn}

The congestion property (CP) means that the more contending
users exist, the less chance a user can grab the channel. Such a property
is natural for practical wireless systems such as the random
backoff and Aloha systems. We can show that the following result (the proof is given in the appendix in the separate supplemental file).
\begin{lem}
\label{lem:If-any-spatial}Assume that any spatial spectrum access game with $N$ users satisfying
the congestion property on a given directed interference graph $G$ has a pure
Nash equilibrium. Then we can construct a new spatial spectrum access game by adding
a new player, whose channel grabbing probability satisfies the congestion property and who has an interference relationship with at most
one player $n\in\mathcal{N}$ in the original game. The new game with $N+1$ users also has a pure Nash
equilibrium.\end{lem}

\begin{defn}[\textbf{Directed Tree} \!\!\cite{Graph}]
A directed graph is called a directed tree if the corresponding undirected
graph obtained by ignoring the directions on the edges of the original
directed graph is a tree.
\end{defn}

Note that a (undirected) tree is a special case of directed trees. Since any spatial spectrum access game over a single node always
has a pure Nash equilibrium, we can then construct the directed tree
recursively by introducing a new node and adding an (directed or undirected)
edge between this node and one existing node. By the induction argument,
we obtain from Lemma \ref{lem:If-any-spatial} that
\begin{cor}\label{thm:tree}
Any spatial spectrum access game satisfying the congestion property
on a directed tree has a pure Nash equilibrium.
\end{cor}
%
%\vspace{-0.05in}

\begin{defn}[\textbf{Directed Forest} \!\!\cite{Graph}]
A directed graph is called a directed forest if it consists of a disjoint union of directed trees.
\end{defn}

Similarly, we can obtain from Lemma \ref{lem:If-any-spatial} that
\begin{cor}
Any spatial spectrum access game satisfying the congestion property
on a directed forest has a pure Nash equilibrium.
\end{cor}

\begin{figure}[tt]
\centering
\includegraphics[scale=0.45]{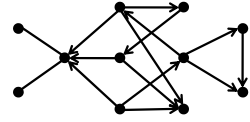}
\caption{\label{fig:A-graph-consists}An interference graph that consists of directed acyclic
graphs and directed trees}
\end{figure}

Note that directed acyclic graph and directed tree structures are widely observed in many wireless networking systems such as wireless ad hoc and sensor networks \cite{santi2005topology}. For example, the star topology with sink/hub nodes in wireless sensor networks is a special case of directed tree. Based on Lemmas \ref{lem:If-any-spatial-1} and \ref{lem:If-any-spatial},
we can construct more general directed interference graphs over which a
spatial spectrum access game satisfying the congestion property has
a pure Nash equilibrium. Figure \ref{fig:A-graph-consists} illustrates such an example.

%This implies that we can study the existence of pure Nash equilibrium through \emph{graph reduction}. Given a directed graph $G$, we can
%obtain the reduced graph $RG$ by removing any node that (1) does not generate interferences to (but can receive interference from) other nodes, or (2) has an interference
%relationship with at most one node.
%According to Lemmas \ref{lem:If-any-spatial-1} and \ref{lem:If-any-spatial}, we know that
%
%\begin{cor}
%If any spatial spectrum game satisfying the congestion property over the reduced graph $RG$ has a pure Nash equilibrium,  then any spatial spectrum game satisfying the congestion property over the original graph $G$  also has a pure Nash equilibrium.
%\end{cor}
%
%By the induction argument, we can then remove the nodes in a similar fashion iteratively  until the reduced graph $RG$ can no longer be simplified. This will significantly simplify the proof  of existence of pure Nash equilibrium of spatial spectrum access games.

\subsection{Existence of Pure Nash Equilibria on Undirected Interference Graphs}

We now study the case that the interference graph is undirected. This is a good approximation of reality if the transmitter of each user is close to its receiver, and all users' interference ranges  are roughly the same. For example, in 802.11 systems, undirected interference graph is widely used to approximate the interference relationships since the access points are typically close to their associated clients and the carrier sensing ranges are similar.

When an undirected interference graph is a tree,  according to Corollary \ref{thm:tree}, any spatial spectrum access game satisfying the congestion property has a pure Nash equilibrium. However, for those non-tree undirected graphs without a topological sort,  the existence of pure Nash equilibrium can not be proved following the results in previous Section \ref{directed}. This motivates us to further study the existence of pure Nash equilibria on generic undirected interference graphs.

First of all, \cite{key99} showed that a $3$-players and $3$-resources congestion
game with user-specific congestion weights may not have a pure Nash equilibrium. Such a congestion game can be considered as a spatial spectrum access game on a complete undirected interference graph (by regarding the resources as channels). When all users have homogeneous channel contention capabilities  and all channels have the same mean data rates, \cite{tekinatomic} showed that the spatial spectrum access game on any undirected interference graphs has a pure Nash equilibrium. Clearly, the applicability of such a channel-homogeneous model is quite limited, since the channel throughputs in practical wireless networks are often heterogeneous. We hence next focus on exploring the random backoff and Aloha systems with user-specific data rates, which provide useful insights for the user-homogeneous and user-heterogeneous channel contention mechanisms, respectively.

Here we  resort to a useful tool of potential game\footnote{Note that it is much more difficult to find a proper potential function to take into account users' asymmetric relationships (i.e., directions of edges on graph) when the interference graph is directed. Hence in this study we only apply the tool of potential game in the undirected case.}, which is defined as
\begin{defn}[\textbf{Potential Game} \!\!\cite{key-41}]
A game is called a potential game if it admits a potential function
$\Phi(\boldsymbol{a})$ such that for every $n\in\mathcal{N}$ and $a_{-n}\in\mathcal{M}^{N-1}$,\begin{align*}
   &\sgn\left(\Phi(a_{n}^{'},a_{-n})-\Phi(a_{n},a_{-n})\right) \\
   = &  \sgn\left(U_{n}(a_{n}^{'},a_{-n})-U_{n}(a_{n},a_{-n})\right),\end{align*}
where $\sgn(\cdot)$ is the sign function.
\end{defn}

%$\Phi(\boldsymbol{a})$ such that for every $n\in\mathcal{N}$ and $a_{-n}\in\mathcal{M}^{N-1}$,\begin{eqnarray*}
% &  & \sgn\left(\Phi(a_{n}^{'},a_{-n})-\Phi(a_{n},a_{-n})\right)\\
% & = & \sgn\left(U_{n}(a_{n}^{'},a_{-n})-U_{n}(a_{n},a_{-n})\right),\end{eqnarray*}

\begin{defn}[\textbf{Better Response Update} \!\!\cite{key-41}]
The event where a player $n$ changes to an action $a_{n}^{'}$ from
the action $a_{n}$ is a better response update if and only if $U_{n}(a_{n}^{'},a_{-n})>U_{n}(a_{n},a_{-n})$.
\end{defn}

An appealing property of the potential game is that it always admits a pure Nash equilibrium and the
finite improvement property, which is defined as

\begin{defn}[\textbf{Finite Improvement Property} \!\!\cite{key-41}]
A game has the finite improvement property if any asynchronous better response
update process (i.e., no more than one player updates the strategy at any given
time) terminates at a pure Nash equilibrium within a finite number of updates.
\end{defn}

Based on the potential game theory, we first study the random backoff mechanism. We show in Theorem \ref{thm:ra1} that when the undirected interference graph is complete, there exists indeed a pure Nash equilibrium.
\begin{thm} \label{thm:ra1}
Any spatial spectrum access game on a complete undirected interference
graph with the random backoff mechanism is a potential game with the potential function \begin{equation}
\Phi(\boldsymbol{a})=\prod_{n=1}^{N}\theta_{a_{n}}B_{a_{n}}^{n}\prod_{m=1}^{M}\prod_{c=0}^{K_{m}(\boldsymbol{a})}\sum_{\lambda=1}^{\lambda_{\max}}\frac{1}{\lambda_{\max}}\left(\frac{\lambda_{\max}-\lambda}{\lambda_{\max}}\right)^{c},\label{eq:pf1}\end{equation} where $K_{m}(\boldsymbol{a})$ is the number of users choosing channel $m$ under the strategy profile $\boldsymbol{a}$, and hence has a pure Nash equilibrium.\end{thm}

The proof is similar to that in \cite{mavronicolas2007congestion}, and hence is omitted here. We then consider the spatial spectrum access game with the random
backoff mechanism on bipartite graphs.
\begin{defn}[\textbf{Bipartite Graph} \!\!\cite{Graph}] An undirected graph is called a bipartite graph if
the set of its nodes can be decomposed into two disjoint sets, such
that no two nodes within the same set are connected by an edge.
\end{defn}

\begin{defn}[\textbf{Complete Bipartite Graph} \!\!\cite{Graph}] An undirected graph is called a complete bipartite
graph if it is a bipartite graph and any two nodes selected from  the
two disjoint sets respectively are connected by an edge.
\end{defn}

\begin{defn}[\textbf{Regular Bipartite Graph} \!\!\cite{Graph}] An undirected graph is called a regular bipartite
graph if it is a bipartite graph and each node is connected by the same number of edges.
\end{defn}

Note that a complete bipartite graph that has the same number of vertices in the two disjoint sets is also a regular bipartite graph. Many well known graphs such as star graphs, circulant graphs, cycle graphs with even number of vertexes, hypercubes, and rectangular lattices are either complete bipartite graphs or regular bipartite graphs. We consider the user specific throughput as
\begin{equation}
U_{n}(a_{n},a_{-n})=h_{n}\theta_{a_{n}}B_{a_{n}}g_{n}(\mathcal{N}_{n}^{a_{n}}(\boldsymbol{a})),
\end{equation}
where $h_{n}$ represents  as a user-specific transmission gain. We can show that
\begin{thm}\label{Thm:bipartite}
Any spatial spectrum access game on either a  complete  bipartite interference
graph or a regular bipartite interference
graph with user specific transmission gains and the random backoff
mechanism has a pure Nash equilibrium.\end{thm}

The proof is given in the appendix in the separate supplemental file. We then consider the random backoff mechanism in the asymptotic case that $\lambda_{\max}$ goes to infinity. This can be a good approximation of reality when the number of backoff mini-slots is much greater than the number of interfering users, and collision rarely occurs. In this case, we have \begin{align}
& g_{n}(\mathcal{N}_{n}^{a_{n}}(\boldsymbol{a}))  = \lim_{\lambda_{\max}\rightarrow\infty}\sum_{\lambda=1}^{\lambda_{\max}}\frac{1}{\lambda_{\max}}\left(\frac{\lambda_{\max}-\lambda}{\lambda_{\max}}\right)^{K_{n}^{a_{n}}(\boldsymbol{a})} \nonumber \\
& = \int_{0}^{1}x^{K_{n}^{a_{n}}(\boldsymbol{a})}dx=\frac{1}{1+K_{n}^{a_{n}}(\boldsymbol{a})},\label{eq:kk}\end{align}
here $K_{n}^{a_{n}}(\boldsymbol{a})$ denotes the number of users that choose channel $a_{n}$ and can interfere with user $n$.  Equation (\ref{eq:kk}) implies that the channel opportunity is equally shared among $1+K_{n}^{a_{n}}(\boldsymbol{a})$ contending users (including user $n$). This can also apply in TDMA channel access mechanism. We consider the user specific throughput as\begin{equation}
U_{n}(a_{n},a_{-n})=h_{n}\theta_{a_{n}}B_{a_{n}}\frac{1}{1+K_{n}^{a_{n}}(\boldsymbol{a})},\label{eq:f11}\end{equation}
We show that
\begin{thm}\label{thm:bb1}
Any spatial spectrum access game on any undirected interference graph with user-specific transmission gains and the random backoff mechanism in the asymptotic case
is a potential game with the potential function
\begin{equation}
\Phi(\boldsymbol{a})=-\sum_{n=1}^{N}\left(\frac{1+\frac{1}{2}K_{n}^{a_{n}}(\boldsymbol{a})}{\theta_{a_{n}}B_{a_{n}}}\right),\label{eq:pf2}\end{equation}
and hence has a pure Nash equilibrium.\end{thm}

Theorem \ref{thm:bb1} is a direct consequence of the more general
result Theorem \ref{thm:bb2}. More specifically, we generalize the throughput function in (\ref{eq:f11}) as\begin{equation}
U_{n}(a_{n},a_{-n})=h_{n}\theta_{a_{n}}B_{a_{n}}\frac{w_{n}}{w_{n}+\sum_{i\in\mathcal{N}_{n}^{a_{n}}(\boldsymbol{a})}w_{i}},\label{eq:gth2}\end{equation}
where $w_{n}>0$ denotes user specific channel sharing weight.  When $w_{n}=1$, the throughput function in (\ref{eq:gth2}) degenerates to the equal-sharing case in (\ref{eq:f11}). The physical meaning
of (\ref{eq:gth2}) is that the channel is shared among the contending
users according to their weights. We refer this to the spatial spectrum access game with user specific sharing weights. Such a proportional channel sharing
scheme has been widely used to model the heterogenous channel access priority assignment for
heterogenous users of different QoS requirements in wireless networks \cite{Monisha2012}. We show that
\begin{thm}\label{thm:bb2}
Any spatial spectrum access game with user specific sharing weights on any undirected interference graph is a potential
game with the potential function as\begin{equation}
\Phi(\boldsymbol{a})=-\sum_{n=1}^{N}\left(\frac{w_{n}^{2}+\frac{1}{2}\sum_{i\in\mathcal{N}_{n}^{a_{n}}(\boldsymbol{a})}w_{n}w_{i}}{\theta_{a_{n}}B_{a_{n}}}\right),\label{eq:ppp3}\end{equation}
and hence has a pure Nash equilibrium.\end{thm}

The proof is given in the appendix in the separate supplemental file. We then consider the random access mechanism under the scenario where
all channels are homogeneous to each user (i.e., $\theta_{m}=\theta$
and $B_{m}=B$ for all $m\in\mathcal{M}$). This models the case where a licensed spectrum band of a large bandwidth (e.g., TV channel) is divided into equal width logical channels for secondary usage with interleaving technology at the physical layer (e.g., the IEEE 802.16d/e standard), such that all channels exhibit the same primary occupancy and have the same channel gain to the same user (but still different channel gains for different users). In this case, the user throughput
function is given as\begin{align}
U_{n}(\boldsymbol{a})=h_{n}\theta Bg(\mathcal{N}_{n}^{a_{n}}(\boldsymbol{a})). \label{eq:good}
\end{align}

We can show that the following result (the proof is given in the appendix in the separate supplemental file).
\begin{thm}\label{Thm:bbbb}
Any spatial spectrum access game on any undirected interference graph
with homogeneous channels, user-specific transmission gains, and the
random backoff mechanism is a potential game with the potential function
\[
\Phi(\boldsymbol{a})=\sum_{n=1}^{N}\left(\frac{1+K_{n}^{a_{n}}(\boldsymbol{a})}{\theta B}\right),
\]
and hence has a pure Nash equilibrium.\end{thm}
%\begin{proof}
%We first define that $\hat{U}_{n}(\boldsymbol{a})=h_{n}\theta BK_{n}^{a_{n}}(\boldsymbol{a})$
%an $\bar{U}_{n}(\boldsymbol{a})=h_{n}\theta B\frac{1}{1+K_{n}^{a_{n}}(\boldsymbol{a})}$.
%It is easy to check that
%\begin{align}
% &\sgn(\hat{U}_{n}(a_{n},a_{-n})-\hat{U}_{n}(a_{n}^{'},a_{-n})) \nonumber \\
%=& -\sgn(\bar{U}_{n}(a_{n},a_{-n})-\bar{U}_{n}(a_{n}^{'},a_{-n})).\label{eq:new1}
%\end{align}
%
%
%Since the random backoff mechanism is adopted, according to (\ref{eq:t1}), we
%have that
%\begin{align}
%&\sgn(U_{n}(a_{n},a_{-n})-U_{n}(a_{n}^{'},a_{-n})) \nonumber \\
%=&\sgn(\hat{U}_{n}(a_{n},a_{-n})-\hat{U}_{n}(a_{n}^{'},a_{-n})).\label{eq:new2}
%\end{align}
%By Theorem \ref{thm:bb1}, we know that the spatial spectrum access game with throughput
%function $\bar{U}_{n}(\boldsymbol{a})$ on any undirected graph is
%a potential game with potential function $\Phi(\boldsymbol{a})=-\sum_{n=1}^{N}\left(\frac{1+K_{n}^{a_{n}}(\boldsymbol{a})}{\theta B}\right)$.
%It hence follows from (\ref{eq:new1}) and (\ref{eq:new2}) that the
%spatial spectrum access game with throughput function $U_{n}(\boldsymbol{a})$ in (\ref{eq:good})
%on any undirected graph is also a potential game with potential function
%$-\Phi(\boldsymbol{a})$. \end{proof}

We now consider the Aloha mechanism. According to (\ref{eq:t2}), we have the throughput function as \begin{equation}
U_{n}(\boldsymbol{a})=\theta_{a_{n}}B_{a_{n}}^{n}p_{n}\prod_{i\in\mathcal{N}_{n}^{a_{n}}(\boldsymbol{a})}(1-p_{i}).\label{eq:pf5}\end{equation}
We can show that the following result (the proof is given in the appendix in the separate supplemental file).
\begin{thm}\label{Aloha}
Any spatial spectrum access game on any undirected interference
graph with the Aloha mechanism is a potential game with the potential function \begin{align*}
\Phi(\boldsymbol{a})  = & \sum_{i=1}^{N}-\log(1-p_{i}) \nonumber \\
& \times \left(\frac{1}{2}\sum_{j\in\mathcal{N}_{i}^{a_{i}}(\boldsymbol{a})}\log(1-p_{j})+\log\left(\theta_{a_{i}}B_{a_{i}}^{i}p_{i}\right)\right),\end{align*}
and hence has a pure Nash equilibrium.
\end{thm}

% As a summary, we depict the results on the existence of pure Nash equilibria of spatial spectrum access games in Figure \ref{fig:Summary}.

\section{Price of Anarchy}\label{POA}
\rev{
In previous sections, we have considered the existence of Nash equilibrium of spatial spectrum access games. We will further explore the efficiency of the Nash equilibrium.

Following the definition of price of anarchy (PoA) in game theory \cite{koutsoupias1999worst}, we will quantify
the efficiency ratio of the worst-case Nash equilibrium over the centralized optimal solution. Let $\Xi$
be the set of Nash equilibria of a given spatial spectrum access game. Then the PoA is defined
as\[
\mbox{PoA}=\frac{\min_{\boldsymbol{a}\in\Xi}\sum_{n\in\mathcal{N}}U_{n}(\boldsymbol{a})}{\max_{\boldsymbol{a}\in\mathcal{M}^{N}}\sum_{n\in\mathcal{N}}U_{n}(\boldsymbol{a})},\]
which is always not greater than $1$. A larger PoA implies that the set of Nash equilibrium is more efficient (in the worst-case sense) using the centralized optimum as a benchmark. Let $V_{n}=\max_{m\in\mathcal{M}} \{\theta_{m}B_{m}^{n}\}$. For a general spatial spectrum access game, we have the following result.

\begin{thm}\label{Proof1:PoA}
The PoA of a spatial spectrum access game $\Gamma=(\mathcal{N},\mathcal{M},G,\{U_{n}\}_{n\in\mathcal{N}})$
is at least $\frac{\min_{n\in\mathcal{N}}\left\{ V_{n}g_{n}(\mathcal{N}_{n})\right\} }{\max_{n\in\mathcal{N}}V_{n}}$. \end{thm}

The proof is given in the appendix in the separate supplemental file. Intuitively, Theorem \ref{Proof1:PoA} indicates that we can increase the efficiency of spectrum sharing by better utilizing the gain of spatial reuse (i.e., reducing the interference edges $\mathcal{N}_{n}$ on the interference graph).

}

\section{Distributed Learning For Spatial Spectrum Access Game}\label{Learning}
As mentioned in Section \ref{game}, both pure and mixed Nash equilibria are important equilibrium concepts for spatial spectrum access games, characterizing system states where  secondary users achieve a mutually satisfactory spectrum sharing solution. We hence consider how to achieve the Nash equilibrium for the spatial spectrum access games in this section.

As shown in Section \ref{NE}, a generic spatial spectrum access game does not necessarily have a pure Nash equilibrium, and thus it is impossible to design a mechanism that is guaranteed to reach a pure Nash equilibrium  in general. However, when the spatial spectrum access game is a potential game, a pure Nash equilibrium exists. In this case, we can apply the Safe Experimentation algorithm in \cite{marden2009payoff} for achieving the pure Nash equilibrium. The key idea is to explore the pure strategy space based on the finite improvement property of the potential game. From the practical application's perspective, we can apply the results obtained in Section \ref{NE} to identify whether the spectrum sharing system satisfies the potential game property. We can then apply the Safe Experimentation algorithm to achieve the pure Nash equilibrium when the  system possesses the potential game property.

Since any spatial spectrum access game always admits a mixed Nash equilibrium, we next target on approaching the mixed Nash equilibria.  From a practical point of view, if the spectrum sharing system is complex and it is difficult to verify the existence of pure Nash equilibrium, we can consider to achieve a mutually satisfactory spectrum sharing solution by allowing users to choose mixed strategies. Govindan and Wilson in \cite{Newton} proposed a global Newton method to compute the mixed Nash equilibria for any finite strategic games. This method hence can be applied to find the mixed Nash equilibria for the spatial spectrum access games. However, such an approach is a centralized optimization, which requires that each user has the complete information of other users and compute the solution accordingly. This is often infeasible in a cognitive radio network, since acquiring complete information requires heavy information exchange among the users, and setting up and maintaining a common control channel for message broadcasting demands high system overheads \cite{key-2}. Moreover, this approach is not incentive compatible since some users may not be willing to share their local information due to the energy consumption of information broadcasting. We thus propose a distributed learning algorithm for any spatial spectrum access games, and the algorithm does not require any information exchange among users. Each user only learns to adjust its channel selection strategy adaptively based on its local throughput observations. We show that the distributed learning algorithm can converge to a mixed Nash equilibrium approximately.

%\begin{figure}[tt]
%\centering
%\includegraphics[scale=0.4]{Period}
%\caption{\label{fig:Period}Time structure of a decision period}
%\end{figure}

\subsection{\label{sec:Distributed-Reinforcement-Learning}Expected Throughput Estimation}

For the distributed learning algorithm, we consider that each user does not have the complete network information and can only estimate its expected throughput locally. Similarly to the approaches in \cite{Chen2012Wiopt} and \cite{key-26}, we can divide the spectrum access time into a sequence of \emph{decision
periods} indexed by $T(=1,2,...)$, where each decision period consists
of $t_{\max}$ time slots. During a single decision period, a user accesses the \emph{same} channel in all $t_{\max}$ time slots in order to better understand the environment.  At the end of
each decision period $T$, a user observes $S_{n}(T,t)$, $I_{n}(T,t)$,
and $b_{n}(T,t)$. Here $S_{n}(T,t)$ denotes the state of the chosen
channel (i.e., whether occupied by the primary traffic), $I_{n}(T,t)$
indicates whether the user has successfully grabbed the channel, i.e.,\[
I_{n}(T,t)=\begin{cases}
1, & \mbox{if user $n$ successfully grabs the channel}\\
0, & \mbox{otherwise,}\end{cases}\]
and $b_{n}(T,t)$ is the received data rate on the chosen channel
by user $n$ at time slot $t$. Note that if $S_{n}(T,t)=0$ (i.e., the channel is occupied by the
primary traffic), we set $I_{n}(T,t)$ and $b_{n}(T,t)$ to be
$0$. At the end of each decision period
$T$, each user $n$ will have a set of local observations $\Omega_{n}(T)=\{S_{n}(T,t),I_{n}(T,t),b_{n}(T,t)\}_{t=1}^{t_{\max}}$. Based on these observations, each user can then apply the Maximum Likelihood Estimation (MLE) method to estimate its expected throughput $U_{n}$. As an example, we next consider the MLE of user expected throughput in the Markovian channel environment introduced in Section \ref{sec:System-Model}.

We first consider the estimation of the channel idle probability $\theta_{m}$.
From the observation set $\Omega_{n}(T)$ at period $T$, user $n$
can observe a sequence of channel state transitions as
\begin{align*}
 &\mathcal{S}_n(T) \\
= & \left(\left(S_{n}(T,1),S_{n}(T,2)\right),...,\left(S_{n}(T,t_{\max}-1),S_{n}(T,t_{\max})\right)\right).
\end{align*}
Here there are four different transition types between adjacent time slots t and t+1: $(0,0),(0,1),(1,0),(1,1)$.
We denote $C_{00}(T),C_{01}(T),C_{10}(T),C_{11}(T)$ as the number
of occurrences of the four state transitions types in $\mathcal{S}_n(T)$,
respectively. According to the principle of MLE, user $n$ can then
compute the likelihood function in terms of channel state transition
parameters $(\varepsilon_{m},\xi_{m})$ as
\begin{align*}
   & \mathcal{L}[\Omega_{n}(T)|\varepsilon_{m},\xi_{m}]=  Pr\{\mathcal{S}_n(T)|\varepsilon_{m},\xi_{m}\}\\
% & = & Pr\{\left(S_{n}(T,1),S_{n}(T,2)\right)|\varepsilon_{m},\xi_{m}\}\\
% & &\times\prod_{t=2}^{t_{\max}-1}Pr\{\left(S_{n}(T,t+1),S_{n}(T,t)\right)|\left(S_{n}(T,1),S_{n}(T,2)\right),\varepsilon_{m},\xi_{m}\}\\
  = & Pr\{S_{n}(T,1)|\varepsilon_{m},\xi_{m}\}
  \\ & \times \prod_{t=1}^{t_{\max}-1}Pr\{S_{n}(T,t+1)|S_{n}(T,t),\varepsilon_{m},\xi_{m}\}\\
  = & Pr\{S_{n}(T,1)|\varepsilon_{m},\xi_{m}\}
  \\ & \times(1-\varepsilon_{m})^{C_{00}(T)}\varepsilon_{m}^{C_{01}(T)}(1-\xi_{m})^{C_{11}(T)}\xi_{m}^{C_{10}(T)}.
\end{align*}
Then MLE of $(\varepsilon_{m},\xi_{m})$ can be computed by maximizing
the log-likelihood function $\ln\mathcal{L}[\Omega_{n}(T)|\varepsilon_{m},\xi_{m}]$,
i.e., $\max_{\varepsilon_{m},\xi_{m}}\ln\mathcal{L}[\Omega_{n}(T)|\varepsilon_{m},\xi_{m}]$.
By the first order condition, we obtain the optimal solution as
\begin{eqnarray*}
\tilde{\varepsilon}_{m} & = & \frac{C_{01}(T)}{C_{00}(T)+C_{01}(T)},\\
\tilde{\xi}_{m} & = & \frac{C_{10}(T)}{C_{11}(T)+C_{10}(T)}.
\end{eqnarray*}
According to (\ref{eq:sd-2}), we can then estimate the channel idle probability
$\theta_{m}$ as
\[
\tilde{\theta}_{m}=\frac{\tilde{\varepsilon}_{m}}{\tilde{\varepsilon}_{m}+\tilde{\xi}_{m}}.
\]

We then consider the estimation of channel grabbing probability $g_n(T)$. When a channel is idle (i.e., no primary traffic), a user $n$ will contend for the channel and can successfully grab the channel with a probability $g_n(T)$. Since there are
a total of $\sum_{t=1}^{t_{\max}}S_{n}(T,t)$ rounds of channel contentions
in the period $T$ and each round is independent, the total number
of successful channel captures $\sum_{t=1}^{t_{\max}}I_{n}(T,t)$ by user $n$ follows
the Binomial distribution. User $n$ then computes the likelihood
of $g_n(T)$ as
\begin{align*}
\mathcal{L}[\Omega_{n}(T)|g_n(T)] = & \left(\begin{array}{c}
\sum_{t=1}^{t_{\max}}S_{n}(T,t)\\
\sum_{t=1}^{t_{\max}}I_{n}(T,t)\end{array}\right)g_m(T)^{\sum_{t=1}^{t_{\max}}I_{n}(T,t)}\nonumber \\
& \times(1-g_m(T))^{\sum_{t=1}^{t_{\max}}S_{n}(T,t)-\sum_{t=1}^{t_{\max}}I_{n}(T,t)}.\label{eq:MLE1}
\end{align*}
Then MLE of $g_n(T)$ can be computed by maximizing the log-likelihood
function $\ln\mathcal{L}[\Omega_{n}(T)|g_n(T)]$, i.e., $\max_{g_n(T)}\ln\mathcal{L}[\Omega_{n}(T)|g_n(T)]$.
By the first order condition, we obtain the optimal solution as \[\tilde{g}_n(T)=\frac{\sum_{t=1}^{t_{\max}}I_{n}(T,t)}{\sum_{t=1}^{t_{\max}}S_{n}(T,t)}.\]

We finally consider the estimation of mean data rate $B_m^n$. Since the received data rate $b_{n}(T,t)$ is also i.i.d. over different
time slots, similar to the MLE
of the channel grabbing probability $g_n(T)$, we can obtain the
MLE of mean data rate $B_{m}^{n}$ as
\[\tilde{B}_{m}^{n}=\frac{\sum_{t=1}^{t_{\max}}b_{n}(T,t)}{\sum_{t=1}^{t_{\max}}I_{n}(T,t)}.\]
By the MLE above, we can then estimate the true expected throughput $U_{n}(T)$
as $\tilde{U}_{n}(T)=\tilde{\theta}_{m}\tilde{B}_{m}^{n}\tilde{g}_n(T).$ In the following analysis of distributed learning algorithm, we consider a general setting where the estimated expected throughput $\tilde{U}_{n}(T)$ of user $n$ can be noisy.  More precisely, we assume that  $\tilde{U}_{n}(T)=U_{n}(T)+w_{n}$ where $w_{n}\in(\underline{w},\overline{w})$
is the random estimation noise with a probability density function
$f_n(w_{n})$ satisfying $E[w_{n}]=\int_{\underline{w}}^{\overline{w}}w_{n}f_n(w_{n})dw_{n}=0$.

\subsection{\label{sec:Distributed-Reinforcement-Learning}Distributed Learning Algorithm}

Based on the expected throughput estimation, we then propose the distributed learning algorithm for spatial spectrum access games.  The idea is to extend the principle of single-agent reinforcement learning to a multi-agent setting. Such multi-agent reinforcement learning algorithm has also been applied to the classical congestion games on complete interference graphs \cite{shah2010dynamics,cominetti2010payoff} by assuming that users are homogeneous (i.e., user's payoff only depends on the number of users choosing the same resource).  Here we extend the learning algorithm to the generalized spatial congestion games on any generic interference graphs with heterogeneous users, which lead to significant differences in analysis. For example, we show that the convergence condition for the learning algorithm depends on the structure of spatial reuse, which is different from those results in \cite{shah2010dynamics,cominetti2010payoff}.

More specifically, at the beginning of each period $T$, a user
$n\in\mathcal{N}$ chooses a channel $a_{n}(T)\in\mathcal{M}$ to
access according to its mixed strategy $\boldsymbol{\sigma}_{n}(T)=(\sigma_{m}^{n}(T),\forall m\in\mathcal{M})$, where $\sigma_{m}^{n}(T)$ is the probability of choosing channel $m$.
The mixed strategy is generated according to $\boldsymbol{P}_{n}(T)=(P_{m}^{n}(T),\forall m\in\mathcal{M})$,
which represents its \emph{perceptions} of the payoff performance of choosing different channels based on local estimations. Perceptions are based on local observations in the past and may not accurately reflect the expected payoff. For example, if a user $n$ has not accessed a channel $m$ for many decision intervals, then perception $P_{m}^{n}(T)$ can be out of date. The key challenge for the learning algorithm is to update the perceptions with proper parameters such that perceptions equal to expected payoffs at the equilibrium.

Similarly to the single-agent  learning, we choose
the Boltzmann distribution as the mapping from perceptions
to mixed strategies, \ie %
\begin{equation}%
\sigma_{m}^{n}(T)=\frac{e^{\gamma P_{m}^{n}(T)}}{\sum_{i=1}^{M}e^{\gamma P_{i}^{n}(T)}},\forall m\in\mathcal{M},\label{eq:MRL-1}%
\end{equation}%
where $\gamma$ is the temperature that controls the randomness
of channel selections. When $\gamma\rightarrow0$, each user will
choose to access channels uniformly at random. When $\gamma\rightarrow\infty$, user $n$ always chooses the channel with the largest perception value $P_{m}^{n}(T)$ among all  channel $m\in\mathcal{M}$. We will show later on that the choice of $\gamma$ trades off convergence and performance of the learning algorithm.

At the end of a decision period $T$, a user $n$ estimated its expected throughput as $\tilde{U}_{n}(\boldsymbol{a}(T))$,  and adjusts its perceptions as\begin{equation}
P_{m}^{n}(T+1)=\begin{cases}
(1-\mu_{T})P_{m}^{n}(T)+\mu_{T}\tilde{U}_{n}(\boldsymbol{a}(T)), & \mbox{if \ensuremath{a_{n}(T)=m,}}\\
P_{m}^{n}(T), & \mbox{otherwise,}\end{cases}\label{eq:MRL-2}\end{equation}
where $\left(\mu_{T}\in(0,1),\forall T\right)$ are the smoothing factors. A user only changes the perception of the channel just accessed in the current decision period, and keeps the perceptions of other channels unchanged.

Algorithm \ref{alg:-Distributed-Reinforcement} summarizes the distributed  learning algorithm. We then analyze the complexity of the distributed  learning algorithm. In each iteration, Line $6$ involves the arithmetic operations over $M$ channels and hence has the complexity of $\mathcal{O}(M)$. The expected throughput estimation in Line $7$ typically involves the arithmetic operations based on the observations of the chosen channel in $t_{\max}$ time slots of the decision period and hence has the complexity of $\mathcal{O}(t_{\max})$. In Line $8$, the perception value update is only carried out in the chosen channel and hence has the complexity of $\mathcal{O}(1)$. Suppose that it takes $K$ iterations for the algorithm to converge. Then total computational complexity of the algorithm is at most $\mathcal{O}(K(M+t_{\max}))$.

\begin{algorithm}[tt]
\begin{algorithmic}[1]
\State \textbf{initialization:}
\State \hspace{0.4cm} \textbf{set} the temperature $\gamma$.
\State \hspace{0.4cm} \textbf{set} the initial perception values $P_{m}^{n}(0)=\frac{1}{M}$ for each user $n\in\mathcal{N}$.
\State \textbf{end initialization \newline}
\Loop{ for each decision period $T$ and each user $n\in\mathcal{N}$ in parallel:}
\State \textbf{select} a channel $m\in\mathcal{M}$ according to (\ref{eq:MRL-1}).
\State \textbf{estimate} the expected throughput $\tilde{U}_{n}(\boldsymbol{a}(T))$.
\State \textbf{update} the perceptions value $\boldsymbol{P}_{n}(T)$ according to (\ref{eq:MRL-2}).
\EndLoop
\end{algorithmic}
\caption{\label{alg:-Distributed-Reinforcement}Distributed Learning Algorithm For Spatial Spectrum Access Game}
\end{algorithm}

\subsection{Convergence of Distributed Learning Algorithm}
We now study the convergence of the proposed distributed learning algorithm based on the theory of stochastic approximation \cite{key-8}.

First, the perception value update in (\ref{eq:MRL-2})
can be written in the following equivalent form, %
\begin{equation}
P_{m}^{n}(T+1)-P_{m}^{n}(T)=\mu_{T}[Z_{m}^{n}(T)-P_{m}^{n}(T)],\forall n\in\mathcal{N},m\in\mathcal{M},\label{eq:CMRL-1}\end{equation}
where $Z_{m}^{n}(T)$ is the update value defined as \begin{equation}
Z_{m}^{n}(T)=\begin{cases}
\tilde{U}_{n}(\boldsymbol{a}(T)), & \mbox{if \ensuremath{a_{n}(T)=m,}}\\
P_{m}^{n}(T), & \mbox{otherwise.}\end{cases}\label{eq:CMRL-1-0}\end{equation}

For the sake of brevity, we denote the perception values, update values, and mixed strategies of all the users as $\boldsymbol{P}(T)\triangleq\left(P_{m}^{n}(T),\forall m\in\mathcal{M}, n\in\mathcal{N}\right)$, $\boldsymbol{Z}(T)\triangleq(Z_{m}^{n}(T),\forall m\in\mathcal{M}, n\in\mathcal{N})$,
and $\boldsymbol{\sigma}(T)\triangleq(\sigma_{m}^{n}(T),\forall m\in\mathcal{M}, n\in\mathcal{N})$, respectively.

Let $Pr\{\mathcal{N}_{n}^{m}(\boldsymbol{a}(T))|\boldsymbol{P}(T),a_{n}(T)=m\}$
denote the conditional probability that, given that the users' perceptions are $\boldsymbol{P}(T)$ and user $n$ chooses channel $m$,  the set of users that choose the same
channel $m$ in user $n$'s neighborhood $\mathcal{N}_{n}$ is $\mathcal{N}_{n}^{m}(\boldsymbol{a}(T))\subseteq\mathcal{N}_{n}$. Since each user independently chooses a
channel according to its mixed strategy $\boldsymbol{\sigma}_{n}(T)$,
then the random set $\mathcal{N}_{n}^{m}(\boldsymbol{a}(T))$ follows the Binomial
distribution of $|\mathcal{N}_{n}|$ independent non-homogeneous Bernoulli
trials with the probability mass function as\begin{align}
    & Pr\{\mathcal{N}_{n}^{m}(\boldsymbol{a}(T))|\boldsymbol{P}(T),a_{n}(T)=m\} \nonumber \\
    = &  \prod_{i\in\mathcal{N}_{n}^{m}(\boldsymbol{a}(T))}(\sigma_{m}^{i}(T))\prod_{i\in\mathcal{N}_{n}\backslash\mathcal{N}_{n}^{m}(\boldsymbol{a}(T))}(1-\sigma_{m}^{i}(T)) \nonumber \\
    = & \prod_{i\in\mathcal{N}_{n}}(\sigma_{m}^{i}(T))^{I_{\{a_{i}(T)=m\}}}(1-\sigma_{m}^{i}(T))^{1-I_{\{a_{i}(T)=m\}}},\label{eq:CDL1}\end{align}
where $I_{\{a_{i}(T)=m\}}=1$ if user $i$ chooses channel $m$, and $I_{\{a_{i}(T)=m\}}=0$ otherwise.

Since the update value $Z_{m}^{n}(T)$ depends on user $n$'s estimated throughput  $\tilde{U}_{n}(\boldsymbol{a}(T))$ (which in turn dependents on $\mathcal{N}_{n}^{m}(\boldsymbol{a}(T))$), thus $Z_{m}^{n}(T)$ is also a random variable. The equations in (\ref{eq:CMRL-1}) are hence stochastic difference equations, which are difficult to analyze directly. We thus focus on the analysis
of its \emph{mean dynamics} \cite{key-8}. To proceed, we define the mapping from the perceptions $\boldsymbol{P}(T)$ to the expected
throughput of user $n$ choosing channel $m$ as $Q_{m}^{n}(\boldsymbol{P}(T))\triangleq E[U_{n}(\boldsymbol{a}(T))|\boldsymbol{P}(T),a_{n}(T)=m]$. Here the expectation $E[\cdot]$ is taken with respective to the mixed strategies $\boldsymbol{\sigma}(T)$ of all users (\ie the perceptions $\boldsymbol{P}(T)$ of all users due to (\ref{eq:MRL-1})). We show that

\begin{lem}\label{lem:l1}
For the distributed learning algorithm, if the temperature satisfies
\begin{align}\gamma<\frac{1}{2\max_{m\in\mathcal{M},n\in\mathcal{N}}\{\theta_{m}B_{m}^{n}\}\max_{n\in\mathcal{N}}\{|\mathcal{N}_{n}|\}},\label{eq:gamma}\end{align}
the mapping from the perceptions to the expected throughput $Q(\boldsymbol{P}(T))\triangleq\left(Q_{m}^{n}(\boldsymbol{P}(T)),m\in\mathcal{M},n\in\mathcal{N}\right)$
forms a maximum-norm contraction.\end{lem}

The proof is given in the appendix in the separate supplemental file. Lemma \ref{lem:l1} implies that when the interference among users becomes more severe (i.e., the maximum degree $\max_{n\in\mathcal{N}}\{|\mathcal{N}_{n}|\}$ of the interference graph becomes larger), a smaller $\gamma$ is needed to guarantee the convergence. This is because that interference relationship among users becomes more complicated and users should put more weight to explore the environment. Note that the condition (\ref{eq:gamma}) is a sufficient condition to form a contraction mapping, which is in turn is a sufficient condition for convergence. Simulation results show that a slightly larger $\gamma$ may also lead to the convergence of the mapping. Based on the property of contraction mapping, there exists
a fixed point $\boldsymbol{P}^{*}$ such that $Q(\boldsymbol{P}^{*})=\boldsymbol{P}^{*}$.
By the theory of stochastic approximations \cite{key-8}, the distributed  learning algorithm will also
converge to the same limit point $\boldsymbol{P}^{*}$.
\begin{thm}\label{thm:Theorem RL-1}
For the distributed learning algorithm, if the temperature $\gamma$ satisfies (\ref{eq:gamma}), $\sum_{T}\mu_{T}=\infty$ and $\sum_{T}\mu_{T}^{2}<\infty$, then
the sequence $\{\boldsymbol{P}(T),\forall T\geq0\}$ converges to
the unique limit point $\boldsymbol{P}^{*}\triangleq(P_{m}^{n*},\forall m\in\mathcal{M},n\in\mathcal{N})$
satisfying that\begin{equation}
Q_{m}^{n}(\boldsymbol{P}^{*})=P_{m}^{n*},\forall m\in\mathcal{M},n\in\mathcal{N}.\label{eq:C2}\end{equation}
\end{thm}

We next explore the property of the equilibrium $\boldsymbol{P}^{*}$
of the distributed learning algorithm. From Theorem \ref{thm:Theorem RL-1}, we see that \begin{equation}
Q_{m}^{n}(\boldsymbol{P}^{*})=E[U_{n}(\boldsymbol{a}(T))|\boldsymbol{P}^{*},a_{n}(T)=m]=P_{m}^{n*}.\label{eq:CRML-2-1}\end{equation}
It means that the perception value $P_{m}^{n*}$ is an accurate estimation
of the expected throughput in the equilibrium. Moreover, we show that
the mixed strategy $\boldsymbol{\sigma}^{*}$ is an approximate Nash equilibrium.
\begin{defn}[\textbf{Approximate Nash Equilibrium} \!\!\cite{key-14}]
A mixed strategy profile $\boldsymbol{\bar{\sigma}}=(\boldsymbol{\bar{\sigma}}_{1},...,\boldsymbol{\bar{\sigma}}_{N})$
is a $\delta$- approximate Nash equilibrium if\[
U_{n}(\boldsymbol{\bar{\sigma}}_{n},\boldsymbol{\bar{\sigma}}_{-n})\ge\max_{\boldsymbol{\sigma}_{n}}U_{n}(\boldsymbol{\sigma}_{n},\boldsymbol{\bar{\sigma}}_{-n})-\delta,\forall n\in\mathcal{N},\]
where $U_{n}(\boldsymbol{\bar{\sigma}}_{n},\boldsymbol{\bar{\sigma}}_{-n})$
denotes the expected throughput of player $n$ under mixed strategy $\boldsymbol{\bar{\sigma}}$,
and $\boldsymbol{\bar{\sigma}}_{-n}$ denotes the mixed strategy profile of other
users except player $n$.
\end{defn}
Here $\delta\geq0$ is the gap from a (precise) mixed Nash equilibrium. For the distributed learning algorithm, we show
that
\begin{thm}
\label{thm:For-the-distributed}For the distributed learning algorithm, the mixed strategy $\boldsymbol{\sigma}^{*}$
in the equilibrium $\boldsymbol{P}^{*}$ is a $\delta$-approximate Nash equilibrium, with
$\delta=\max_{n\in\mathcal{N}}\{-\frac{1}{\gamma}\sum_{m=1}^{M}\sigma_{m}^{n*}\ln\sigma_{m}^{n*}\}$.\end{thm}

The proof is given in the appendix in the separate supplemental file. The gap $\delta$ can be interpreted as the \emph{weighted entropy}, which describes the randomness of the learning exploration. A larger $\delta$ means a worse learning performance.  When each user adopts the uniformly random access, the gap $\delta$ reaches the maximum value and results in the worst learning performance. In this case, we can obtain the upper-bound of the gap $\delta$ as $-\frac{1}{\gamma}\sum_{m=1}^{M}\frac{1}{M}\ln\frac{1}{M}=\frac{1}{\gamma}\ln M.$ Theorems \ref{thm:Theorem RL-1} and \ref{thm:For-the-distributed} together illustrate the trade-off between the exploration and exploitation through the choice of $\gamma$.
A small enough $\gamma$ is required to explore the environment (so that users will not put too many weights on exploitation and get stuck in channels with the \emph{current} best throughputs) and guarantee the convergence
of distributed  learning to the approximate mixed Nash equilibrium. If $\gamma$ is too small, however, then the performance gap $\delta$ is large due to over-exploration. Numerical results in Section \ref{sec:Numerical-Results} demonstrate that, by a proper choice of the temperature $\gamma$, the performance loss of the approximate mixed Nash equilibrium obtained  by distributed learning is at most $10\%$ compared with the exact mixed Nash equilibrium.

\section{Extension to Physical Interference Model}
\rev{
In previous sections, we have focused on studying the spatial spectrum access game under the protocol interference model, which has been widely adopted in wireless network research literature \cite{shi2009correctly,zhou2013practical}. The protocol interference model uses the interference graph to describe the pair-wide interference relationships among users, i.e., two users can interfere with each other if they are within each other's interference range. This is useful for modeling the data transmission confliction when some random access MAC protocol is adopted.   For example, in 802.11 networks, we can construct the interference graph by setting the carrier sensing range as the interference range. Moreover, by carefully constructing the interference edges via the reality check approach in \cite{shi2009correctly} or the measurement-calibrated propagation scheme in \cite{zhou2013practical}, the protocol interference model can provide a good approximation to the physical interference model that captures the continuous nature of interference and takes into account the accumulated interference from multiple concurrent transmitters \cite{gupta2000capacity,yang2008physical}.

We next study the spatial spectrum access game under the setting of physical interference model. According to \cite{gupta2000capacity}, we can compute the data rate of user $n$ under the physical interference model as
\begin{equation}
U_{n}(\boldsymbol{a})=\theta_{m}W\log_{2}\left(1+\frac{\eta_{n}d_{n}^{-\alpha}}{\omega_{0}+\omega_{a_{n}}^{n}+\sum_{i\in\mathcal{N}/\{n\}:a_{i}=a_{n}}\eta_{i}d_{in}^{-\alpha}}\right).\label{eq:U}
\end{equation}
Here $W$ is the channel bandwidth, $\eta_{i}$ is the transmission
power of user $n$, $\alpha$ is the path loss factor, and $d_{n}^{-\alpha}$
denotes the channel gain between the transmitter and the receiver
of user $n$ due to free-space attenuation \cite{gupta2000capacity}. Furthermore,
$\omega_{0}$ denotes the background noise, $\omega_{a_{n}}^{n}$
denotes the the interference from primary users to secondary user
$n$ on  channel $a_{n}$, $\eta_{i}d_{in}^{-\alpha}$ denotes
the interference generated by user $i$ to user $n$, and $\sum_{i\in\mathcal{N}\backslash\{n\}:a_{i}=a_{n}}\eta_{i}d_{in}^{-\alpha}$
denotes the accumulated interference from other second users
to user $n$. Similarly to the setting of protocol interference model, we can model the distributed spectrum access
problem among users under the physical interference model as a spatial
spectrum access game with the payoff function given as in (\ref{eq:U}).
We can show the following result for the case that the channel availability $\theta_{m}$ of all channels are homogeneous, i.e., $\theta_{m}=\theta$ for any $m\in\mathcal{M}$.
\begin{thm}
\label{thm:SINR}When the channel availabilities of all channels are homogeneous, the spatial spectrum access game
under the physical interference model is a potential game
with the potential function given as
\begin{align*}
\Phi(\boldsymbol{a})= -\sum_{i}\sum_{j\neq i}\eta_{i}\eta_{j}d_{ij}^{-\alpha}I_{\{a_{i}=a_{j}\}}-2\sum_{n=1}^{N}\eta_{n}\left(\omega_{a_{n}}^{n}+\omega_{0}\right),
\end{align*}
 and hence has a pure Nash equilibrium.\end{thm}

The proof is given in the appendix in the separate supplemental file. Since the spatial spectrum access game is a potential game, we can then apply the Safe Experimentation algorithm in \cite{marden2009payoff} for achieving the pure Nash equilibrium. For the general case that the channel availabilities are heterogeneous, the analysis  of spatial spectrum access game is very challenging and will be considered in a future work.
} 
\section{Numerical Results}\label{sec:Numerical-Results}

We now evaluate the proposed distributed learning algorithm by simulations.
We consider a Rayleigh fading channel environment. The data rate of
user $n$ on an idle channel $m$ is given according to the Shannon
capacity, i.e., $b_{m}^{n}(t)=W\log_{2}\left(1+\frac{\eta_{n}z_{m}^{n}(t)}{\omega_{m}^{n}}\right)$.
We consider the Rayleigh fading
channel environment where channel gain $z_{m}^{n}(t)$ is a random variable that follows the exponential
distribution with the mean $\bar{z}_{m}^{n}$. In the following simulations,
we set $W=10$ MHz, $\omega_{m}^{n}=-100$ dBm, and $\eta_{n}=100$ mW.
By choosing different mean channel gains $\bar{z}_{m}^{n}$, we have
different mean data rates $B_{m}^{n}=E[b_{m}^{n}(t)]$ for different
channels and users. We set the channel idle probability $\theta_{m}=0.5$.

\begin{figure}[tt]
\centering
\includegraphics[scale=0.45]{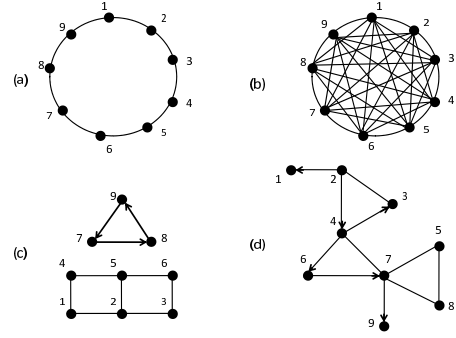}
\caption{\label{fig:Interference-Graphs}Interference Graphs}
\end{figure}

We consider a network of $M=5$ channels and $N=9$ users with four
different interference graphs (see Figure \ref{fig:Interference-Graphs}).
Graphs (a) and (b) are undirected, and Graphs (c) and (d) are directed.
Let $\boldsymbol{B}_{n}=\{B_{1}^{n},...,B_{M}^{n}\}$ be the mean data rate
vector of user $n$. We set $\boldsymbol{B}_{1}=\boldsymbol{B}_{2}=\boldsymbol{B}_{3}=\{2,6,16,20,30\}$
Mbps, $\boldsymbol{B}_{4}=\boldsymbol{B}_{5}=\boldsymbol{B}_{6}=\{4,12,32,40,60\}$ Mbps,
and $\boldsymbol{B}_{7}=\boldsymbol{B}_{8}=\boldsymbol{B}_{9}=\{10,30,80,100,150\}$ Mbps.
We implement both the random backoff and Aloha mechanisms for channel contention.
For the random backoff mechanism, we set the number of backoff mini-slots
in a time slot $\lambda_{\max}=10$. For the Aloha mechanism, the
channel contention probabilities of the users are randomly assigned from the set $\{0.3,0.5,0.7\}$.  Notice that in this study we focus on channel choices instead of the adjustment of contention probabilities. For the distributed learning algorithm initialization,we set the smooth factor $\mu_{T}=\frac{1}{T}$,
which satisfies the condition $\sum_{T}\mu_{T}=\infty$ and $\sum_{T}\mu_{T}^{2}<\infty$.

%\begin{figure}[tt]
%\centering
%\includegraphics[scale=0.37]{system}
%\caption{\label{fig:The-system-performance}The system performance of the distributed
%learning algorithm with different temperature $\gamma$}
%\end{figure}

%\begin{figure}[tt]
%\begin{minipage}[t]{0.48\linewidth}
%\centering
%\includegraphics[scale=0.45]{InGraph}
%\caption{\label{fig:Interference-Graphs}Interference Graphs}
%\end{minipage}
%\hfill
%\begin{minipage}[t]{0.48\linewidth}
%\centering
%\includegraphics[scale=0.37]{system}
%\caption{\label{fig:The-system-performance}The system performance of the distributed
%learning algorithm with different temperature $\gamma$}
%\end{minipage}
%\end{figure}

\begin{figure*}[tt]
\begin{minipage}[t]{0.48\linewidth}
\centering
\includegraphics[scale=0.4]{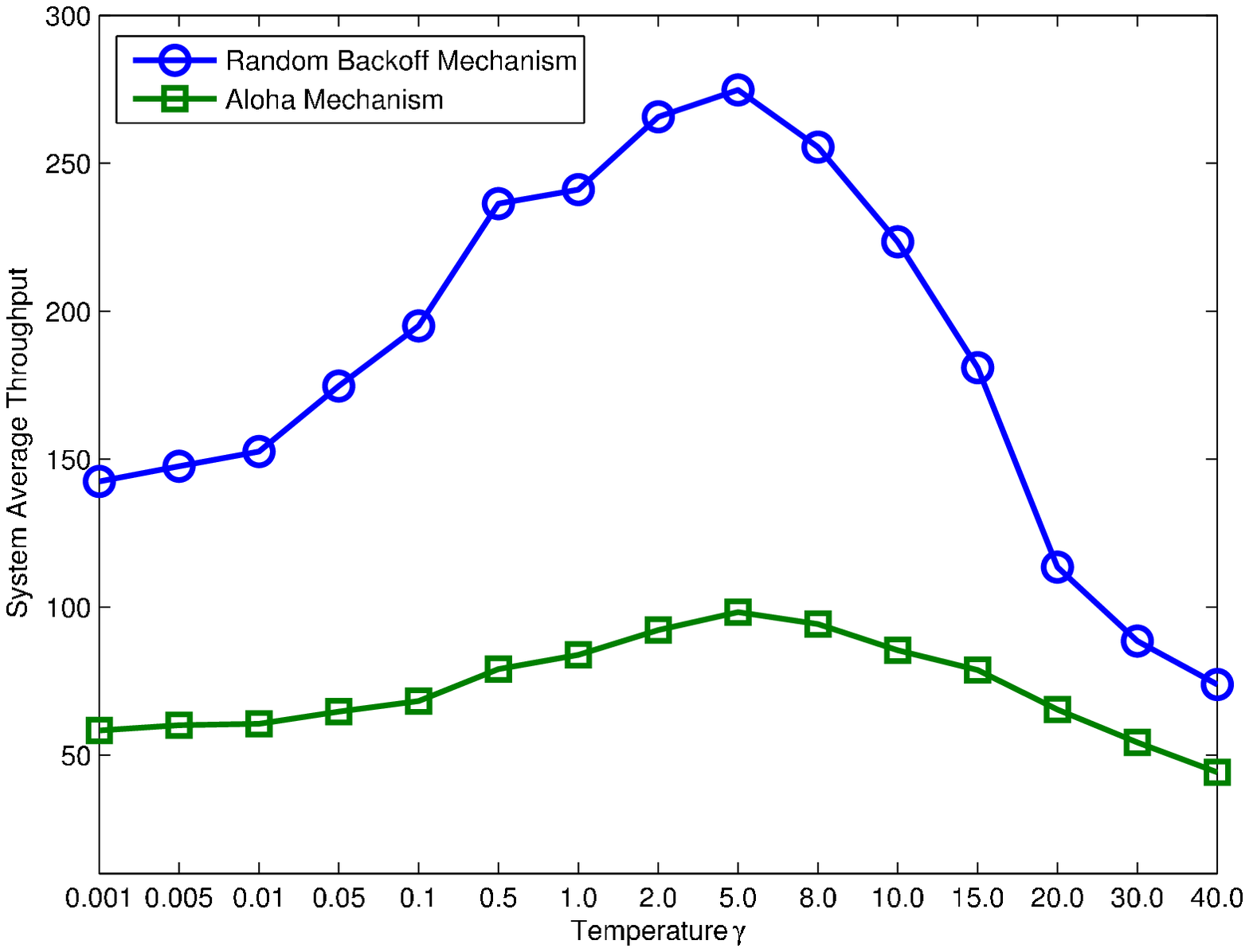}
\caption{\label{fig:The-system-performance}The system performance of the distributed
learning algorithm with different temperature $\gamma$}
\end{minipage}
\hfill
\begin{minipage}[t]{0.48\linewidth}
\centering
\includegraphics[scale=0.48]{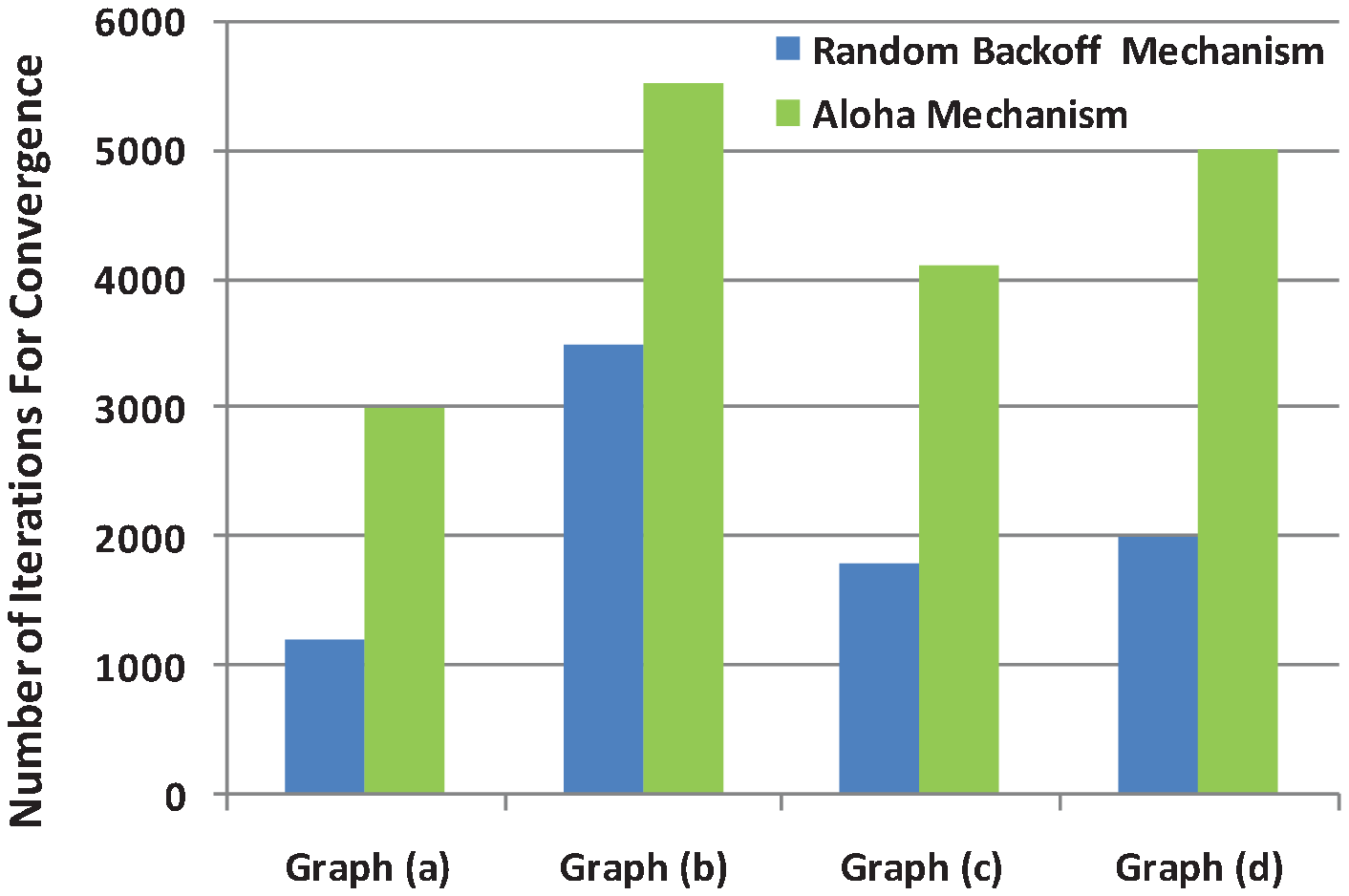}
\caption{\label{fig:Convergence}Convergence time of distributed learning algorithm on different graphs}
\end{minipage}
\end{figure*}

We first evaluate the distributed learning algorithm
with different choices of temperature $\gamma$ on the interference
graph (d) in Figure \ref{fig:Interference-Graphs}. We run the learning
algorithm sufficiently long until the time average system throughput does
not change. The result in Figure \ref{fig:The-system-performance} shows the system performance with different $\gamma$, and demonstrates
that a proper temperature $\gamma$ can achieve a balance between exploration and exploitation and offer the best performance.
When is $\gamma$ small, the users tend to select the channels randomly (i.e., over-exploration) and the performance gap $\delta$ can be large. When $\gamma$
is very large, the algorithm focuses on exploitation and may get stuck in local optimum and the performance
is again negatively affected. In the following
simulations we set $\gamma=5.0$ since it achieves good system performance in both random
backoff and Aloha mechanisms as in Figure \ref{fig:The-system-performance}.

%\begin{figure}[tt]
%\begin{minipage}[t]{0.48\linewidth}
%\centering
%\includegraphics[scale=0.37]{system}
%\caption{\label{fig:The-system-performance}The system performance of the distributed
%learning algorithm with different temperature $\gamma$}
%\end{minipage}
%\hfill
%\begin{minipage}[t]{0.48\linewidth}
%\centering
%\includegraphics[scale=0.45]{Convegernce2}
%\caption{\label{fig:Convegernce2}The convergence time of the distributed
%learning algorithm with different temperature $\gamma$}
%\end{minipage}
%\end{figure}

We then evaluate the convergence of the distributed learning algorithm. In Figure \ref{fig:Convergence}, we show the number of iterations for the convergence of distributed learning algorithm with random backoff and Aloha mechanisms. We see that, as the interference graph becomes more dense (e.g., graph (b)), the convergence time becomes longer. The reason is that, when the interference graph becomes more dense and a user can generate interference to more users, the environment becomes more complex and it hence takes more time overhead to explore. We also observe that the convergence time of Aloha mechanism is longer than that of  random backoff mechanism. This is because that in Aloha mechanism users are heterogeneous in terms of channel contention capability and hence the system environment becomes more complicated.

%\begin{figure*}[tt]
%\begin{minipage}[t]{0.48\linewidth}
%\centering
%\includegraphics[scale=0.4]{fig3}
%\caption{\label{fig:Users'-average-throughput-1}Users' average throughput
%on interference graph (d) with Aloha mechanism}
%\end{minipage}
%\hfill
%\begin{minipage}[t]{0.48\linewidth}
%\centering
%\includegraphics[scale=0.4]{fig1}
%\caption{\label{fig:Users'-average-throughput}Users' average throughput on
%interference graph (d) with random backoff mechanism}
%\end{minipage}
%\end{figure*}

\begin{figure*}[tt]
\begin{minipage}[t]{0.48\linewidth}
\centering
\includegraphics[scale=1.2]{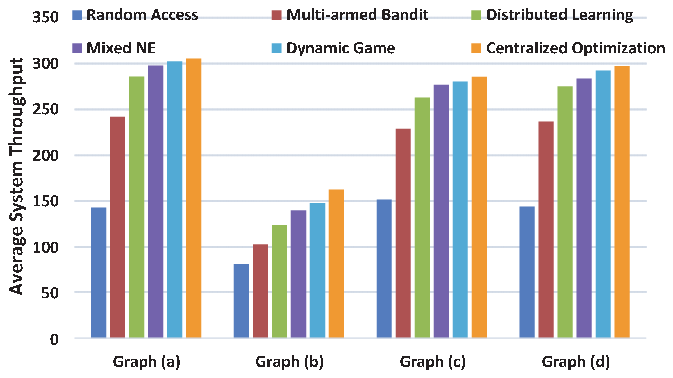}
\caption{\label{fig:Comparison-of-distributed}Comparison of distributed learning,
random access, global Newton, and centralized optimization with the random backoff mechanism}
\end{minipage}
\hfill
\begin{minipage}[t]{0.48\linewidth}
\centering
\includegraphics[scale=1.2]{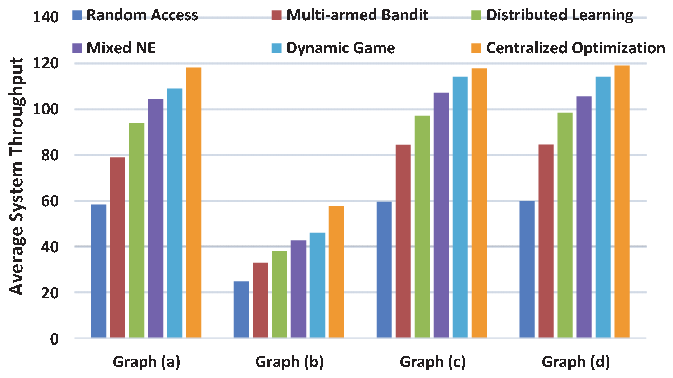}
\caption{\label{fig:Comparison-of-distributed-1}Comparison of distributed
learning, random access, global Newton, and centralized optimization with the Aloha mechanism}
\end{minipage}
\end{figure*}

We further compare distributed learning with the solutions obtained by the following algorithms:
\begin{itemize}
\item Random Access: each user chooses a channel to access purely randomly.
\item Multi-armed Bandit: we implement the multi-armed bandit solution for distributed spectrum access in \cite{lai2011cognitive}.
\item Mixed NE: we compute the exact mixed Nash equilibrium of the spatial spectrum access game, based on statistical network information using the global Newton method \cite{Newton}.
\item Dynamic Game: we compute the solution of the complete information dynamic game for spectrum access, by solving the corresponding stage spatial spectrum access game (using the global Newton method) for each time slot $t$, with the channel idle probabilities $\{\theta_m\}_{m=1}^{M}$ of the stage game replaced by the channel state realization $\{S_m(t)\}_{m=1}^{M}$ of time slot $t$.
\item Centralized Optimization: the solution obtained by solving the centralized
global optimization of $\\ \max_{\boldsymbol{a}}\sum_{n\in\mathcal{N}}U_{n}(\boldsymbol{a})$.
\end{itemize}

We implement these algorithms together with the distributed learning
algorithm on the four types of interference graphs in Figure \ref{fig:Interference-Graphs}.
The results are shown in Figures \ref{fig:Comparison-of-distributed}
and \ref{fig:Comparison-of-distributed-1}. We see that the exact mixed Nash equilibrium of spatial spectrum access game based on statistical network information is efficient, with a less than $8\%$ performance loss compared with the dynamic game solution. This is because that in the proposed spatial spectrum access game, each secondary user takes both the primary activity levels on different channels and the competition with other secondary users into consideration. This can help to mitigate the congestion within secondary users and exploit the channels of high transmission opportunities.  Moreover, the performance gap $\delta$ of the approximate mixed Nash equilibrium by distributed learning from the exact mixed Nash equilibrium is small, with a less than $10\%$ performance loss in all cases.  For the random backoff
(Aloha, respectively) mechanism, we see that the distributed learning
algorithm achieves up-to $100\%$ ($65\%$, respectively) performance
improvement over the random access algorithm. Moreover, we observe that the distributed learning algorithm can achieve better performance than the multi-armed bandit scheme, with a performance gain of up-to $15\%$. This is because that the design of multi-armed bandit scheme does not take the spatial reuse effect in account. Compared with the centralized optimal solution, the performance loss of the distributed
learning in the full-interference graph (b) is $28\%$ ($34\%$, respectively). Such performance loss is not due to the algorithm design; instead it is due to the selfish nature of the users (i.e., competition in distributed spectrum sharing).
In the partial-interference graphs (a), (c), and (d), the performance
loss can be further reduced to less than $10\%$ ($17\%$, respectively). This shows that the negative impact of users' selfish behavior is smaller when users can share the spectrum more efficiently through spatial reuse.

\section{Conclusion}\label{sec:Conclusion}
In this paper, we explore the spatial aspect of distributed spectrum sharing, and propose a framework of spatial spectrum access game on directed interference graphs. We investigate the critical issue of the existence of pure Nash equilibria, and develop a distributed learning algorithm converging to an approximate mixed Nash equilibrium for any spatial spectrum access games. Numerical results show that the algorithm is
efficient and has significant performance gain over a random access algorithm that does not take the spatial effect into consideration.

For the future work, we are going to design distributed spectrum access mechanisms that can converge to pure Nash equilibria for those spatial spectrum games which are not potential games but admit pure Nash equilibria. We will also study the general spatial spectrum access game under the physical interference model.

%\section*{ACKNOWLEDGMENTS}
%This work is supported by the General Research Funds (Project Number 412509,  412710, and 412511) established under the University Grant Committee of the Hong Kong Special Administrative Region, China.

\bibliographystyle{ieeetran}
\bibliography{DynamicSpectrum}

\newpage
\section{Appendix}\label{appendixA}

\subsection{\label{proof111} Proof of Lemma \ref{lem:If-any-spatial-1}} 
Suppose that the original spatial spectrum access game on the interference
graph $G$ has a pure Nash equilibrium $\boldsymbol{a}^{*}.$ We index the newly
added player as player $N+1$, denote the set of players that can
generate interference to player $N+1$ as $\mathcal{I}(\boldsymbol{a}^{*})$, and denote
the set of player $N+1$'s interfering players that choose
channel $m$ as $\mathcal{I}_{m}(\boldsymbol{a}^{*})\triangleq\{n\in\mathcal{I}(\boldsymbol{a}^{*}):a_{n}^{*}=m\}$.
Now player $N+1$ can compute its best response strategy $a_{N+1}^{*}=\arg\max_{a\in\mathcal{M}}\{\theta_{a}B_{a}^{N+1}g_{N+1}(\mathcal{I}_{a}(\boldsymbol{a}^{*}))\}$.
Obviously, the strategy profile $(\boldsymbol{a}^{*},a_{N+1}^{*})$ of $N+1$ players
in the new game is a pure Nash equilibrium. \qed

\subsection{\label{proof112} Proof of Lemma \ref{lem:If-any-spatial}} 
Suppose that the original spatial spectrum access game has a pure Nash equilibrium $\boldsymbol{a}^{*}.$ We index the newly
added player as player $N+1$. Now player $N+1$ can compute its best
response strategy $a_{N+1}^{*}$ as follows.

Case $1$: if player $N+1$ does not have an interference relationship with any
player in the original game, then \begin{eqnarray}a_{N+1}^{*}=\arg\max_{a\in\mathcal{M}}\{\theta_{a}B_{a}^{N+1}g_{N+1}(\varnothing)\}.\label{eq:d1}\end{eqnarray}
In this case, the strategy profile $(\boldsymbol{a}^{*},a_{N+1}^{*})$ of $N+1$
players in the new game is a pure Nash equilibrium.

Case $2$: if player $N+1$ can not generate interference to any player in the
original game, and there is one player $n\in\mathcal{N}$ that can
generate interference to player $N+1$, then \begin{align}
& a_{N+1}^{*} \nonumber \\
= &  \arg\max_{a\in\mathcal{M}}\{\max_{a\neq a_{n}^{*}}\{\theta_{a}B_{a}^{N+1}g_{N+1}(\varnothing)\},\theta_{a_{n}^{*}}B_{a_{n}^{*}}^{N+1}g_{N+1}(\{n\})\}.\label{eq:d2}\end{align}
According to Lemma \ref{lem:If-any-spatial-1}, the strategy profile
$(\boldsymbol{a}^{*},a_{N+1}^{*})$ must be a pure Nash equilibrium of the new
game.

Case $3$: If player $N+1$ can generate interference to a player $n\in\mathcal{N}$
in the original game, then player $N+1$ can compute its best response strategy $a_{N+1}^{*}$ according to (\ref{eq:d2}) if player $n$ can also generate interference to player $N+1$, and
according to (\ref{eq:d1}) otherwise. By checking that whether $a_{N+1}^{*}$ and $a_{n}^{*}$ are the same, we have the following two possibilities.

1) If $a_{N+1}^{*}\neq a_{n}^{*}$, then the strategy profile $(\boldsymbol{a}^{*},a_{N+1}^{*})$
must be a pure Nash equilibrium for the new game.

2) If $a_{N+1}^{*}=a_{n}^{*}$, we modify the original game as follows.
For player $n$, we modify the utility function as $\hat{U}_{n}(\boldsymbol{a})=\theta_{a_{n}}B_{a_{n}}^{n}g_{n}(\mathcal{N}_{n}^{a_{n}}(\boldsymbol{a})\cup\{N+1\})$
if $a_{n}=a_{N+1}^{*}$ and $\hat{U}_{n}(\boldsymbol{a})=U_{n}(\boldsymbol{a})$ otherwise.
For any other player $i\neq n$ and $i\in\mathcal{N}$, we keep the
utility function unchanged, i.e., $\hat{U}_{i}(\boldsymbol{a})=U_{i}(\boldsymbol{a}).$ For
the modified game with the utility function $\{\hat{U}_{i}(\boldsymbol{a})\}_{i\in\mathcal{N}}$,
it is easy to check that it also satisfies the congestion property and hence it has a pure Nash equilibrium $\hat{\boldsymbol{a}}^{*}$ by assumption (since this game is still played over the same given interference graph and any game satisfying the congestion property on original interference graph has a pure Nash equilibrium by assumption). If
$\hat{a}_{n}^{*}=a_{N+1}^{*}$, the strategy profile $(\hat{\boldsymbol{a}}^{*},a_{N+1}^{*})$
is a pure Nash equilibrium for the new game with $N+1$ players, since
given the interference between player $n$ and player $N+1$, any
player $i\in\mathcal{N}\cup\{N+1\}$ is playing the best response
strategy. If $\hat{a}_{n}^{*}\neq a_{N+1}^{*}$, we can show that the strategy profile
$(\hat{\boldsymbol{a}}^{*},a_{N+1}^{*})$ is still a pure Nash equilibrium for
the new game. This is because that, without the interference between
player $n$ and player $N+1$ (since they play different strategies),
playing strategy $a_{N+1}^{*}$ can not be worse due to the congestion
property $g_{N+1}(\varnothing)\geq g_{N+1}(\{n\})$ and hence is still player
$N+1$'s best response. For any player $i\in\mathcal{N}$ in the original game, playing
$\hat{a}_{i}^{*}$ is the best response strategy given player $N+1$'s
strategy $a_{N+1}^{*}$. \qed

\subsection{\label{proof1122} Proof of Theorem \ref{Thm:bipartite}} 
We first consider the complete bipartite graph. Since the interference graph is bipartite, we can
decompose the set of vertices into two disjoint sets $\mathcal{V}_{1}$
and $\mathcal{V}_{2}$. Let $D_{i}=|\mathcal{V}_{i}|$ denote the
number of vertices in set $\mathcal{V}_{i}$ for $i\in\{1,2\}$. Without
loss of generality, we will assume that $D_{1}\geq D_{2}$. According
to the definition of bipartite graph, we see that two users
can generate interference to each other if and only they are from
different sets.

Recall that the channel grabbing probability of the random backoff mechanism is given in  (\ref{eq:t1}). We define that
$f(K_{n}^{a_{n}}(\boldsymbol{a}))\triangleq g_{n}(\mathcal{N}_{n}^{a_{n}}(\boldsymbol{a}))$, where $K_{n}^{a_{n}}(\boldsymbol{a})$ is the number of contending
users with user $n$. We then order the channels such that $\theta_{\tilde{1}}B_{\tilde{1}}f(0)\geq\theta_{\tilde{2}}B_{\tilde{2}}f(0)\geq...\geq\theta_{\tilde{M}}B_{\tilde{M}}f(0)$. If $\theta_{\tilde{1}}B_{\tilde{1}}f(D_{1})\geq\theta_{\tilde{2}}B_{\tilde{2}}f(0)$,
we can trivially construct a pure Nash equilibrium by assigning each
user with channel $\tilde{1}$. If $\theta_{\tilde{1}}B_{\tilde{1}}f(D_{1})<\theta_{\tilde{2}}B_{\tilde{2}}f(0),$
we can construct a pure Nash equilibrium by assigning users in set
$\mathcal{V}_{1}$ with channel $\tilde{1}$ and users in set $\mathcal{V}_{2}$
with channel $\tilde{2}$. In this case, the throughput of a user
$n$ in set $\mathcal{V}_{1}$ is $h_{n}\theta_{\tilde{1}}B_{\tilde{1}}f(0)\geq h_{n}\theta_{\tilde{2}}B_{\tilde{2}}f(D_{2})$
and the throughput of a user $n^{'}$ in set $\mathcal{V}_{2}$ is
$h_{n^{'}}\theta_{\tilde{2}}B_{\tilde{2}}f(0)>h_{n^{'}}\theta_{\tilde{1}}B_{\tilde{1}}f(D_{1}).$
No user has  the incentive to deviate.

For the regular bipartite graph case, we define that $D_{1}=D_{2}=D$ where $D$ is the node degree of the graph (i.e., the number of edges connected by each node). It is then easy to check that the above procedure of constructing pure Nash equilibrium is still valid.   \qed

\subsection{\label{proof1} Proof of Theorem \ref{thm:bb2}}
For the ease of exposition, we first define that\[
\Phi_{n}^{m}(\boldsymbol{a})=-\left(\frac{\frac{1}{2}\sum_{i\in\mathcal{N}_{n}^{a_{n}}(\boldsymbol{a})}w_{n}w_{i}}{\theta_{a_{n}}B_{a_{n}}}\right)I_{\{a_{n}=m\}}.\]
Thus, we have $\Phi(\boldsymbol{a})=\sum_{n=1}^{N}\sum_{m=1}^{M}\Phi_{n}^{m}(\boldsymbol{a})-\sum_{n=1}^{N}\frac{w_{n}^{2}}{\theta_{a_{k}}B_{a_{k}}}$.

Now suppose that a user $k$ unilaterally changes its strategy $a_{k}$
to $a_{k}^{'}$, such that the strategy profile changes from $\boldsymbol{a}$
to $\boldsymbol{a}^{'}$. We have that\begin{align}
   & \Phi(\boldsymbol{a}^{'})-\Phi(\boldsymbol{a})\nonumber \\
  = & \sum_{m=1}^{M}\Phi_{k}^{m}(\boldsymbol{a}^{'})-\sum_{m=1}^{M}\Phi_{k}^{m}(\boldsymbol{a})+\sum_{i\in\mathcal{N}_{k}}\left(\Phi_{i}^{a_{k}^{'}}(\boldsymbol{a}^{'})-\Phi_{i}^{a_{k}^{'}}(\boldsymbol{a})\right) \nonumber \\ & +\sum_{i\in\mathcal{N}_{k}}\left(\Phi_{i}^{a_{k}}(\boldsymbol{a}^{'})-\Phi_{i}^{a_{k}}(\boldsymbol{a})\right)-\frac{w_{k}^{2}}{\theta_{a_{k}^{'}}B_{a_{k}^{'}}}+\frac{w_{k}^{2}}{\theta_{a_{k}}B_{a_{k}}}.\label{eq:hhhh1}\end{align}
For the part $\sum_{m=1}^{M}\Phi_{k}^{m}(\boldsymbol{a}^{'})-\sum_{m=1}^{M}\Phi_{k}^{m}(\boldsymbol{a})$,
we have\begin{align}
& \sum_{m=1}^{M}\Phi_{k}^{m}(\boldsymbol{a}^{'})-\sum_{m=1}^{M}\Phi_{k}^{m}(\boldsymbol{a})= \Phi_{k}^{a_{k}^{'}}(\boldsymbol{a}^{'})-\Phi_{k}^{a_{k}}(\boldsymbol{a}) \nonumber \\
& = -\frac{w_{k}\frac{1}{2}\sum_{i\in\mathcal{N}_{k}^{a_{k}^{'}}(\boldsymbol{a}^{'})}w_{i}}{\theta_{a_{k}^{'}}B_{a_{k}^{'}}}+\frac{w_{k}\frac{1}{2}\sum_{i\in\mathcal{N}_{k}^{a_{k}}(\boldsymbol{a})}w_{i}}{\theta_{a_{k}}B_{a_{k}}}.\label{eq:hhhh3}\end{align}
For the part $\Phi_{i}^{a_{k}^{'}}(\boldsymbol{a}^{'})-\Phi_{i}^{a_{k}^{'}}(\boldsymbol{a})$,
we have\begin{align*}
 & \Phi_{i}^{a_{k}^{'}}(\boldsymbol{a}^{'})-\Phi_{i}^{a_{k}^{'}}(\boldsymbol{a}) \\
  =  & -\frac{\frac{1}{2}w_{i}}{\theta_{a_{k}^{'}}B_{a_{k}^{'}}}\left(\sum_{j\in\mathcal{N}_{i}^{a_{k}^{'}}(\boldsymbol{a}^{'})}w_{j}\right)I_{\{a_{i}=a_{k}^{'}\}}
   \\ & +\frac{\frac{1}{2}w_{i}}{\theta_{a_{k}^{'}}B_{a_{k}^{'}}}\left(\sum_{j\in\mathcal{N}_{i}^{a_{k}^{'}}(\boldsymbol{a})}w_{j}\right)I_{\{a_{i}=a_{k}^{'}\}}\\
 = & -\frac{\frac{1}{2}w_{i}}{\theta_{a_{k}^{'}}B_{a_{k}^{'}}}\left(\sum_{j\in\mathcal{N}_{i}^{a_{k}^{'}}(\boldsymbol{a}^{'})}w_{j}-\sum_{j\in\mathcal{N}_{i}^{a_{k}^{'}}(\boldsymbol{a})}w_{j}\right)I_{\{a_{i}=a_{k}^{'}\}}
 \\ = & -\frac{\frac{1}{2}w_{i}w_{k}}{\theta_{a_{k}^{'}}B_{a_{k}^{'}}}I_{\{a_{i}=a_{k}^{'}\}}.\end{align*}
This implies that\begin{align}
& \sum_{i\in\mathcal{N}_{k}}\left(\Phi_{i}^{a_{k}^{'}}(\boldsymbol{a}^{'})-\Phi_{i}^{a_{k}^{'}}(\boldsymbol{a})\right)=\sum_{i\in\mathcal{N}_{k}}-\frac{\frac{1}{2}w_{i}w_{k}}{\theta_{a_{k}^{'}}B_{a_{k}^{'}}}I_{\{a_{i}=a_{k}^{'}\}}
\nonumber \\
= & -\frac{\frac{1}{2}w_{k}}{\theta_{a_{k}^{'}}B_{a_{k}^{'}}}\sum_{i\in\mathcal{N}_{k}^{a_{k}^{'}}(\boldsymbol{a}^{'})}w_{i}.\label{eq:hhhh4}\end{align}
Similarly, we have that\begin{align}
 \sum_{i\in\mathcal{N}_{k}}\left(\Phi_{i}^{a_{k}}(\boldsymbol{a}^{'})-\Phi_{i}^{a_{k}}(\boldsymbol{a})\right)= \frac{\frac{1}{2}w_{k}}{\theta_{a_{k}}B_{a_{k}}}\sum_{i\in\mathcal{N}_{k}^{a_{k}}(\boldsymbol{a})}w_{i}.\label{eq:hhhh5}\end{align}

Substituting (\ref{eq:hhhh3}), (\ref{eq:hhhh4}), and (\ref{eq:hhhh5})
into (\ref{eq:hhhh1}), we obtain that\begin{align*}
   & \Phi(\boldsymbol{a}^{'})-\Phi(\boldsymbol{a}) \\
  = & -\frac{w_{k}\sum_{i\in\mathcal{N}_{k}^{a_{k}^{'}}(\boldsymbol{a}^{'})}w_{i}}{\theta_{a_{k}^{'}}B_{a_{k}^{'}}}-\frac{w_{k}^{2}}{\theta_{a_{k}^{'}}B_{a_{k}^{'}}}+\frac{w_{k}\sum_{i\in\mathcal{N}_{k}^{a_{k}}(\boldsymbol{a})}w_{i}}{\theta_{a_{k}}B_{a_{k}}}+\frac{w_{k}^{2}}{\theta_{a_{k}}B_{a_{k}}}\\
   = & \frac{1}{h_{k}}\theta_{a_{k}^{'}}B_{a_{k}^{'}}\theta_{a_{k}}B_{a_{k}}\sum_{i\in\mathcal{N}_{k}^{a_{k}^{'}}(\boldsymbol{a}^{'})}w_{i}\sum_{i\in\mathcal{N}_{k}^{a_{k}}(\boldsymbol{a})}w_{i}\left(U_{k}(\boldsymbol{a}^{'})-U_{k}(\boldsymbol{a})\right),\end{align*}
which completes the proof. \qed

\subsection{\label{proof233444} Proof of Theorem \ref{Thm:bbbb}}
We first define that $\hat{U}_{n}(\boldsymbol{a})=h_{n}\theta BK_{n}^{a_{n}}(\boldsymbol{a})$
an $\bar{U}_{n}(\boldsymbol{a})=h_{n}\theta B\frac{1}{1+K_{n}^{a_{n}}(\boldsymbol{a})}$.
It is easy to check that
\begin{align}
 &\sgn(\hat{U}_{n}(a_{n},a_{-n})-\hat{U}_{n}(a_{n}^{'},a_{-n})) \nonumber \\
=& -\sgn(\bar{U}_{n}(a_{n},a_{-n})-\bar{U}_{n}(a_{n}^{'},a_{-n})).\label{eq:new1}
\end{align}

Since the random backoff mechanism is adopted, according to (\ref{eq:t1}), we
have that
\begin{align}
&\sgn(U_{n}(a_{n},a_{-n})-U_{n}(a_{n}^{'},a_{-n})) \nonumber \\
=&\sgn(\hat{U}_{n}(a_{n},a_{-n})-\hat{U}_{n}(a_{n}^{'},a_{-n})).\label{eq:new2}
\end{align}
By Theorem \ref{thm:bb1}, we know that the spatial spectrum access game with throughput
function $\bar{U}_{n}(\boldsymbol{a})$ on any undirected graph is
a potential game with potential function $\Phi(\boldsymbol{a})=-\sum_{n=1}^{N}\left(\frac{1+K_{n}^{a_{n}}(\boldsymbol{a})}{\theta B}\right)$.
It hence follows from (\ref{eq:new1}) and (\ref{eq:new2}) that the
spatial spectrum access game with throughput function $U_{n}(\boldsymbol{a})$ in (\ref{eq:good})
on any undirected graph is also a potential game with potential function
$-\Phi(\boldsymbol{a})$. \qed

\subsection{\label{proof2} Proof of Theorem \ref{Aloha}}
For the ease of exposition, we first define $\rho_{i}\triangleq\log(1-p_{i})$,
$\xi_{m,d}^{i}\triangleq\log(\theta_{m}B_{m,d}^{i}p_{i})$, and\[
\Phi_{i}^{m}(\boldsymbol{a})=-\rho_{i}\left(\frac{1}{2}\sum_{j\in\mathcal{N}_{i}^{m}(\boldsymbol{a})}\rho_{j}+\xi_{m,d_{i}}^{i}\right)I_{\{a_{i}=m\}}.\]
Thus, we have $\Phi(\boldsymbol{a})=\sum_{i=1}^{N}\sum_{m=1}^{M}\Phi_{i}^{m}(\boldsymbol{a})$.

Now suppose that a user $k$ unilaterally changes its strategy $a_{k}$
to $a_{k}^{'}$. Let $\boldsymbol{a}'=(a_{1},...,a_{k-1},a_{k}^{'},a_{k+1},...,a_{N})$
be the new strategy profile. Thus, the change in potential $\Phi$
from $\boldsymbol{a}$ to $\boldsymbol{a}^{'}$ is given by\begin{align}
 & \Phi(\boldsymbol{a}^{'})-\Phi(\boldsymbol{a})= \sum_{i=1}^{N}\sum_{m=1}^{M}\Phi_{i}^{m}(\boldsymbol{a}^{'})-\sum_{i=1}^{N}\sum_{m=1}^{M}\Phi_{i}^{m}(\boldsymbol{a})\nonumber \\
= & \sum_{m=1}^{M}\Phi_{k}^{m}(\boldsymbol{a}^{'})-\sum_{m=1}^{M}\Phi_{k}^{m}(\boldsymbol{a})+\sum_{i\in\mathcal{N}_{k}}\sum_{m=1}^{M}\Phi_{i}^{m}(\boldsymbol{a}^{'})-\sum_{i\in\mathcal{N}_{k}}\sum_{m=1}^{M}\Phi_{i}^{m}(\boldsymbol{a})\nonumber \\
= & \left(\sum_{m=1}^{M}\Phi_{k}^{m}(\boldsymbol{a}^{'})-\sum_{m=1}^{M}\Phi_{k}^{m}(\boldsymbol{a})\right)+\sum_{i\in\mathcal{N}_{k}}\left(\Phi_{i}^{a_{k}^{'}}(\boldsymbol{a}^{'})-\Phi_{i}^{a_{k}^{'}}(\boldsymbol{a})\right) \nonumber\\
& +\sum_{i\in\mathcal{N}_{k}}\left(\Phi_{i}^{a_{k}}(\boldsymbol{a}^{'})-\Phi_{i}^{a_{k}}(\boldsymbol{a})\right).\label{eq:P1}\end{align}

Equation (\ref{eq:P1}) consists of three parts. Next we analyze each part separately. For the first part, we have\begin{align}
 & \sum_{m=1}^{M}\Phi_{k}^{m}(\boldsymbol{a}^{'})-\sum_{m=1}^{M}\Phi_{k}^{m}(\boldsymbol{a})= \Phi_{k}^{a_{k}^{'}}(\boldsymbol{a}^{'})-\Phi_{k}^{a_{k}}(\boldsymbol{a}) \nonumber \\
 = &  -\rho_{k}\left(\frac{1}{2}\sum_{j\in\mathcal{N}_{k}^{a_{k}^{'}}(\boldsymbol{a}^{'})}\rho_{j}+\xi_{a_{k}^{'},d_{k}}^{k}\right)+\rho_{k}\left(\frac{1}{2}\sum_{j\in\mathcal{N}_{k}^{a_{k}}(\boldsymbol{a})}\rho_{j}+\xi_{a_{k},d_{k}}^{k}\right).\label{eq:P2}\end{align}

For the second part in (\ref{eq:P1}), \begin{align*}
 & \Phi_{i}^{a_{k}^{'}}(\boldsymbol{a}^{'})-\Phi_{i}^{a_{k}^{'}}(\boldsymbol{a}) \\
 = & -\rho_{i}\left(\frac{1}{2}\sum_{j\in\mathcal{N}_{i}^{a_{k}^{'}}(\boldsymbol{a}^{'})}\rho_{j}+\xi_{a_{k}^{'},d_{i}}^{i}\right)I_{\{a_{i}=a_{k}^{'}\}}
\\ & +\rho_{i}\left(\frac{1}{2}\sum_{j\in\mathcal{N}_{i}^{a_{k}^{'}}(\boldsymbol{a})}\rho_{j}+\xi_{a_{k}^{'},d_{i}}^{i}\right)I_{\{a_{i}=a_{k}^{'}\}}\\
= & -\frac{1}{2}\rho_{i}\left(\sum_{j\in\mathcal{N}_{i}^{a_{k}^{'}}(\boldsymbol{a}^{'})}\rho_{j}-\sum_{j\in\mathcal{N}_{i}^{a_{k}^{'}}(\boldsymbol{a})}\rho_{j}\right)I_{\{a_{i}=a_{k}^{'}\}}
\\ = & -\frac{1}{2}\rho_{i}\rho_{k}I_{\{a_{i}=a_{k}^{'}\}}.\end{align*}
This means \begin{align}
 & \sum_{i\in\mathcal{N}_{k}}\left(\Phi_{i}^{a_{k}^{'}}(\boldsymbol{a}^{'})-\Phi_{i}^{a_{k}^{'}}(\boldsymbol{a})\right) \nonumber \\ = & \sum_{i\in\mathcal{N}_{k}}-\frac{1}{2}\rho_{i}\rho_{k}I_{\{a_{i}=a_{k}^{'}\}}=-\frac{1}{2}\rho_{k}\sum_{i\in\mathcal{N}_{k}^{a_{k}^{'}}(\boldsymbol{a}^{'})}\rho_{i}.\label{eq:P4}\end{align}
For the third term in (\ref{eq:P1}), we can similarly get\begin{align}
 & \sum_{i\in\mathcal{N}_{k}}\left(\Phi_{i}^{a_{k}}(\boldsymbol{a}^{'})-\Phi_{i}^{a_{k}}(\boldsymbol{a})\right)=\frac{1}{2}\rho_{k}\sum_{i\in\mathcal{N}_{k}^{a_{k}^{'}}(\boldsymbol{a})}\rho_{i}.\label{eq:P5-1}\end{align}
Substituting (\ref{eq:P2}), (\ref{eq:P4}), and (\ref{eq:P5-1}) into
(\ref{eq:P1}), we obtain\begin{align}
 & \Phi(\boldsymbol{a}^{'})-\Phi(\boldsymbol{a}) \nonumber \\
 = & -\rho_{k}\left(\sum_{j\in\mathcal{N}_{k}^{a_{k}^{'}}(\boldsymbol{a}^{'})}\rho_{j}+\xi_{a_{k}^{'},d_{k}}^{k}-\sum_{j\in\mathcal{N}_{k}^{a_{k}}(\boldsymbol{a})}\rho_{j}-\xi_{a_{k},d_{k}}^{k}\right) \nonumber \\
 = & -\rho_{k}\left(\log U_{k}(\boldsymbol{a}^{'})-\log U_{k}(\boldsymbol{a})\right).\label{eq:P6}\end{align}

Since $-\rho_{k}=-\log(1-p_{k})>0$ and \begin{align*} & \sgn\left(U_{n}(a_{n}^{'},a_{-n})-U_{n}(a_{n},a_{-n})\right)\\
= &  \sgn\left(\log U_{n}(a_{n}^{'},a_{-n})-\log U_{n}(a_{n},a_{-n})\right),\end{align*} we can conclude
that $\Phi(\boldsymbol{a})$  is a
potential function. \qed

\subsection{\label{proof3456} Proof of Theorem \ref{Proof1:PoA}}
First of all, since $g_{n}(\mathcal{N}_{n}^{a_{n}}(\boldsymbol{a}))\leq1$,
we know that $U_{n}(\boldsymbol{a})=\theta_{m}B_{m}^{n}g_{n}(\mathcal{N}_{n}^{a_{n}}(\boldsymbol{a}))\leq V_{n}$.
Thus, it follows that
\begin{equation}
\max_{\boldsymbol{a}}\sum_{n\in\mathcal{N}}U_{n}(\boldsymbol{a})\leq\sum_{n\in\mathcal{N}}V_{n}.\label{eq:PoA1}
\end{equation}
Suppose that $\widetilde{\boldsymbol{a}}\in\Xi$ is an arbitrary Nash
equilibrium of the spatial spectrum access game $\Gamma$. Then at
Nash equilibrium, we must have that
\begin{equation}
U_{n}(\widetilde{\boldsymbol{a}})\geq V_{n}g_{n}(\mathcal{N}_{n}).\label{eq:PoA2}
\end{equation}
Otherwise, the user $n$ can always improve its payoff by choosing
the channel that maximizes $\theta_{m}B_{m}^{n}$, which would contradict
with the fact that $\widetilde{\boldsymbol{a}}$ is a Nash equilibrium.
According to (\ref{eq:PoA1}) and (\ref{eq:PoA2}), we then obtain
that
\begin{eqnarray*}
\mbox{PoA} & \geq & \frac{\sum_{n\in\mathcal{N}}U_{n}(\widetilde{\boldsymbol{a}})}{\max_{\boldsymbol{a}}\sum_{n\in\mathcal{N}}U_{n}(\boldsymbol{a})}\\
 & \geq & \frac{\sum_{n\in\mathcal{N}}V_{n}g_{n}(\mathcal{N}_{n})}{\sum_{n\in\mathcal{N}}V_{n}}\\
 & \geq & \frac{N\min_{n\in\mathcal{N}}\left\{ V_{n}g_{n}(\mathcal{N}_{n})\right\} }{N\max_{n\in\mathcal{N}}V_{n}}\\
 & = & \frac{\min_{n\in\mathcal{N}}\left\{ V_{n}g_{n}(\mathcal{N}_{n})\right\} }{\max_{n\in\mathcal{N}}V_{n}}.
\end{eqnarray*} \qed

\subsection{\label{proof3} Proof of Lemma \ref{lem:l1}}
This proof is derived based on the one in \cite{cominetti2010payoff}. The key difference is that we consider that the generalized spatial congestion games on any generic graphs and  users can be heterogenous. While, \cite{cominetti2010payoff} only considers the special case that the congestion game is defined over the complete graphs and users are homogenous.

According to (\ref{eq:CDL1}), we first compute the expected payoff as\begin{align}
   & Q_{m}^{n}(\boldsymbol{P}(T))= \sum_{\mathcal{N}_{n}^{m}(\boldsymbol{a}(T))\subseteq\mathcal{N}_{n}}\theta_{m}B_{m}^{n} \nonumber \\
  & \times g_{n}(\mathcal{N}_{n}^{m}(\boldsymbol{a}(T)))Pr\{\mathcal{N}_{n}^{m}(\boldsymbol{a}(T))|\boldsymbol{P}(T),a_{n}(T)=m\}.\label{eq:CDL2-1}\end{align}

We now consider the difference in expected payoffs between $Q_{m}^{n}(\boldsymbol{P}(T))$
and $Q_{m}^{n}(\hat{\boldsymbol{P}}(T))$ given two arbitrary perceptions
$\boldsymbol{P}(T)$ and $\hat{\boldsymbol{P}}(T)$ as
\begin{align}
   & |Q_{m}^{n}(\boldsymbol{P}(T))-Q_{m}^{n}(\boldsymbol{\hat{P}}(T))|\nonumber \\
  = & \theta_{m}B_{m}^{n}|\sum_{\mathcal{N}_{n}^{m}(\boldsymbol{a}(T))\subseteq\mathcal{N}_{n}}g_{n}(\mathcal{N}_{n}^{m}(\boldsymbol{a}(T)))\nonumber \\
   & \times\prod_{i\in\mathcal{N}_{n}}(\sigma_{m}^{i}(T))^{I_{\{a_{i}(T)=m\}}}(1-\sigma_{m}^{i}(T))^{1-I_{\{a_{i}(T)=m\}}}\nonumber \\
   & -\sum_{\mathcal{N}_{n}^{m}(\boldsymbol{a}(T))\subseteq\mathcal{N}_{n}}g_{n}(\mathcal{N}_{n}^{m}(\boldsymbol{a}(T)))\nonumber \\
   & \times\prod_{i\in\mathcal{N}_{n}}(\hat{\sigma}_{m}^{i}(T))^{I_{\{a_{i}(T)=m\}}}(1-\hat{\sigma}_{m}^{i}(T))^{1-I_{\{a_{i}(T)=m\}}}|\nonumber \\
  \leq & \theta_{m}B_{m}^{n}\sum_{\mathcal{N}_{n}^{m}(\boldsymbol{a}(T))\subseteq\mathcal{N}_{n}}g_{n}(\mathcal{N}_{n}^{m}(\boldsymbol{a}(T)))\nonumber \\
   & \times|\prod_{i\in\mathcal{N}_{n}}(\sigma_{m}^{i}(T))^{I_{\{a_{i}(T)=m\}}}(1-\sigma_{m}^{i}(T))^{1-I_{\{a_{i}(T)=m\}}}\nonumber \\
   & -\prod_{i\in\mathcal{N}_{n}}(\hat{\sigma}_{m}^{i}(T))^{I_{\{a_{i}(T)=m\}}}(1-\hat{\sigma}_{m}^{i}(T))^{1-I_{\{a_{i}(T)=m\}}}|\nonumber \\
  \leq & \theta_{m}B_{m}^{n}\sum_{\mathcal{N}_{n}^{m}(\boldsymbol{a}(T))\subseteq\mathcal{N}_{n}} \nonumber \\
  & |\prod_{i\in\mathcal{N}_{n}}(\sigma_{m}^{i}(T))^{I_{\{a_{i}(T)=m\}}}(1-\sigma_{m}^{i}(T))^{1-I_{\{a_{i}(T)=m\}}}\nonumber \\
   & -\prod_{i\in\mathcal{N}_{n}}(\hat{\sigma}_{m}^{i}(T))^{I_{\{a_{i}(T)=m\}}}(1-\hat{\sigma}_{m}^{i}(T))^{1-I_{\{a_{i}(T)=m\}}}|.\label{eq:lls1}\end{align}
For sake of brevity, we define that \begin{align*}
   & \psi(\boldsymbol{\sigma}(T),\hat{\boldsymbol{\sigma}}(T),k)\\
  \triangleq & \prod_{i\leq k:i\in\mathcal{N}_{n}}(\sigma_{m}^{i}(T))^{I_{\{a_{i}(T)=m\}}}(1-\sigma_{m}^{i}(T))^{1-I_{\{a_{i}(T)=m\}}}\\
   & \times\prod_{i>k:i\in\mathcal{N}_{n}}(\hat{\sigma}_{m}^{i}(T))^{I_{\{a_{i}(T)=m\}}}(1-\hat{\sigma}_{m}^{i}(T))^{1-I_{\{a_{i}(T)=m\}}}.\end{align*}
Here we index all users in the set $\mathcal{N}_{n}$ arbitrarily
as $1,2,...,|\mathcal{N}_{n}|$. Obviously, we have that \begin{align*}
   & |\psi(\boldsymbol{\sigma}(T),\hat{\boldsymbol{\sigma}}(T),k)-\psi(\boldsymbol{\sigma}(T),\hat{\boldsymbol{\sigma}}(T),k-1)|\\
  = & \prod_{i<k:i\in\mathcal{N}_{n}}(\sigma_{m}^{i}(T))^{I_{\{a_{i}(T)=m\}}}(1-\sigma_{m}^{i}(T))^{1-I_{\{a_{i}(T)=m\}}}\\
   & \times\prod_{i>k:i\in\mathcal{N}_{n}}(\hat{\sigma}_{m}^{i}(T))^{I_{\{a_{i}(T)=m\}}}(1-\hat{\sigma}_{m}^{i}(T))^{1-I_{\{a_{i}(T)=m\}}}\\
   & \times|\sigma_{m}^{k}(T)-\hat{\sigma}_{m}^{k}(T)|,\forall k=1,...,|\mathcal{N}_{n}|.\end{align*}
Since $\prod_{i<k:i\in\mathcal{N}_{n}}(\sigma_{m}^{k}(T)+1-\sigma_{m}^{k}(T))\prod_{i>k:i\in\mathcal{N}_{n}}(\hat{\sigma}_{m}^{i}(T)+1-\hat{\sigma}_{m}^{i}(T))=1$,
we obtain that\begin{align}
   & \sum_{\mathcal{N}_{n}^{m}(\boldsymbol{a}(T))\subseteq\mathcal{N}_{n}}|\psi(\boldsymbol{\sigma}(T),\hat{\boldsymbol{\sigma}}(T),k)-\psi(\boldsymbol{\sigma}(T),\hat{\boldsymbol{\sigma}}(T),k-1)|\nonumber \\
  = & \sum_{\mathcal{N}_{n}^{m}(\boldsymbol{a}(T))\subseteq\mathcal{N}_{n}:a_{k}(T)=m}|\psi(\boldsymbol{\sigma}(T),\hat{\boldsymbol{\sigma}}(T),k)-\psi(\boldsymbol{\sigma}(T),\hat{\boldsymbol{\sigma}}(T),k-1)|\nonumber \\
   +&\sum_{\mathcal{N}_{n}^{m}(\boldsymbol{a}(T))\subseteq\mathcal{N}_{n}:a_{k}(T)\neq m}|\psi(\boldsymbol{\sigma}(T),\hat{\boldsymbol{\sigma}}(T),k)-\psi(\boldsymbol{\sigma}(T),\hat{\boldsymbol{\sigma}}(T),k-1)|\nonumber \\
  =& 2\sum_{\mathcal{N}_{n}^{m}(\boldsymbol{a}(T))\subseteq\mathcal{N}_{n}:a_{k}(T)=m}|\sigma_{m}^{k}(T)-\hat{\sigma}_{m}^{k}(T)|\nonumber \\
   & \times\prod_{i<k:i\in\mathcal{N}_{n}}(\sigma_{m}^{i}(T))^{I_{\{a_{i}(T)=m\}}}(1-\sigma_{m}^{i}(T))^{1-I_{\{a_{i}(T)=m\}}}\nonumber \\
   & \times\prod_{i>k:i\in\mathcal{N}_{n}}(\hat{\sigma}_{m}^{i}(T))^{I_{\{a_{i}(T)=m\}}}(1-\hat{\sigma}_{m}^{i}(T))^{1-I_{\{a_{i}(T)=m\}}}\nonumber \\
  =& 2|\sigma_{m}^{k}(T)-\hat{\sigma}_{m}^{k}(T)|.\label{eq:DSL55}\end{align}
Then, we have from (\ref{eq:lls1}) and (\ref{eq:DSL55}) that\begin{align}
   & |Q_{m}^{n}(\boldsymbol{P}(T))-Q_{m}^{n}(\boldsymbol{\hat{P}}(T))|\nonumber \\
  \leq & \theta_{m}B_{m}^{n} \nonumber \\
  & \times\sum_{\mathcal{N}_{n}^{m}(\boldsymbol{a}(T))\subseteq\mathcal{N}_{n}}|\psi(\boldsymbol{\sigma}(T),\hat{\boldsymbol{\sigma}}(T),|\mathcal{N}_{n}|)-\psi(\boldsymbol{\sigma}(T),\hat{\boldsymbol{\sigma}}(T),0)|\nonumber \\
  = & \theta_{m}B_{m}^{n}\sum_{\mathcal{N}_{n}^{m}(\boldsymbol{a}(T))\subseteq\mathcal{N}_{n}}|\sum_{k=0}^{|\mathcal{N}_{n}|-1}\left(\psi(\boldsymbol{\sigma}(T),\hat{\boldsymbol{\sigma}}(T),|\mathcal{N}_{n}|-k)\right.\nonumber \\
   & \left.-\psi(\boldsymbol{\sigma}(T),\hat{\boldsymbol{\sigma}}(T),|\mathcal{N}_{n}|-k-1)\right)|\nonumber \\
  \leq & \theta_{m}B_{m}^{n}\sum_{\mathcal{N}_{n}^{m}(\boldsymbol{a}(T))\subseteq\mathcal{N}_{n}}\sum_{k=0}^{|\mathcal{N}_{n}|-1}|\psi(\boldsymbol{\sigma}(T),\hat{\boldsymbol{\sigma}}(T),|\mathcal{N}_{n}|-k)\nonumber \\
   & -\psi(\boldsymbol{\sigma}(T),\hat{\boldsymbol{\sigma}}(T),|\mathcal{N}_{n}|-k-1)|\nonumber \\
  = & 2\theta_{m}B_{m}^{n}\sum_{k=1}^{|\mathcal{N}_{n}|}|\sigma_{m}^{k}(T)-\hat{\sigma}_{m}^{k}(T)|.\label{eq:CDL3}\end{align}

We then define a function $f(\boldsymbol{P}_{i}(T))\triangleq\sigma_{m}^{i}(T)=\frac{e^{\gamma P_{m}^{i}(T)}}{\sum_{m'=1}^{M}e^{\gamma P_{m'}^{i}(T)}}$.
Since $f(\boldsymbol{P}_{i}(T))$ is continuously differentiable,
by the mean value theorem, we know that there exists $\boldsymbol{\bar{P}}_{i}(T)=\delta\left(\boldsymbol{P}_{i}(T)-\boldsymbol{\hat{P}}_{i}(T)\right)$
with $0<\delta<1$ such that\begin{align*}
   & \sigma_{m}^{i}(T)-\hat{\sigma}_{m}^{i}(T)\\
  = & \frac{e^{\gamma P_{m}^{i}(T)}}{\sum_{m'=1}^{M}e^{\gamma P_{m'}^{i}(T)}}-\frac{e^{\gamma\hat{P}_{m}^{i}(T)}}{\sum_{m'=1}^{M}e^{\gamma\hat{P}_{m'}^{i}(T)}}\\
  = & \gamma\left[\frac{e^{\gamma\bar{P}_{m}^{i}(T)}\sum_{m'=1}^{M}e^{\gamma\bar{P}_{m'}^{i}(T)}-e^{2\gamma\bar{P}_{m}^{i}(T)}}{\left(\sum_{m'=1}^{M}e^{\gamma\bar{P}_{m'}^{i}(T)}\right)^{2}}\right]\left(P_{m}^{i}(T)-\hat{P}_{m}^{i}(T)\right)\\
   & -\sum_{m'=1,m'\neq m}^{M}\gamma\frac{e^{\gamma\bar{P}_{m'}^{i}(T)}e^{\gamma\bar{P}_{m}^{i}(T)}}{\left(\sum_{m'=1}^{M}e^{\gamma\bar{P}_{m'}^{i}(T)}\right)^{2}}\left(P_{m}^{i}(T)-\hat{P}_{m}^{i}(T)\right).\end{align*}
Let $C_{m}^{i}=\frac{e^{\gamma\bar{P}_{m}^{i}(T)}\sum_{m'=1}^{M}e^{\gamma\bar{P}_{m'}^{i}(T)}-e^{2\gamma\bar{P}_{m}^{i}(T)}}{\left(\sum_{m'=1}^{M}e^{\gamma\bar{P}_{m'}^{i}(T)}\right)^{2}}$
and $C_{m'}^{i}=\frac{e^{\gamma\bar{P}_{m'}^{i}(T)}e^{\gamma\bar{P}_{m}^{i}(T)}}{\left(\sum_{m'=1}^{M}e^{\gamma\bar{P}_{m'}^{i}(T)}\right)^{2}}$.
It is easy to check that $C_{m}^{i}=\sum_{m'=1,m'\neq m}^{M}C_{m'}^{i}$
and $2C_{m}^{i}\leq1$. Thus,\begin{align}
   & |\sigma_{m}^{i}(T)-\hat{\sigma}_{m}^{i}(T)|\nonumber \\
  \leq & \gamma C_{m}^{i}|P_{m}^{i}(T)-\hat{P}_{m}^{i}(T)|+\sum_{m'=1,m'\neq m}^{M}C_{m'}^{i}|P_{m}^{i}(T)-\hat{P}_{m}^{i}(T)|\nonumber \\
  \leq & \gamma\left(C_{m}^{i}+\sum_{m'=1,m'\neq m}^{M}C_{m'}^{i}\right)||\boldsymbol{P}_{i}(T)-\hat{\boldsymbol{P}}_{i}(T)||_{\infty}\nonumber \\
  \leq & \gamma||\boldsymbol{P}_{i}(T)-\hat{\boldsymbol{P}}_{i}(T)||_{\infty}.\label{eq:CDL5}\end{align}
Combining (\ref{eq:CDL5}) and (\ref{eq:CDL3}), we obtain\begin{align*}
   & |Q_{m}^{n}(\boldsymbol{P}(T))-Q_{m}^{n}(\hat{\boldsymbol{P}}(T))|\\
  \leq & 2\gamma\theta_{m}B_{m}^{n}\sum_{i\in\mathcal{N}_{n}}||\boldsymbol{P}_{i}(T)-\hat{\boldsymbol{P}}_{i}(T)||_{\infty}\\
  \leq & 2\gamma\theta_{m}B_{m}^{n}|\mathcal{N}_{n}|||\boldsymbol{P}(T)-\hat{\boldsymbol{P}}(T)||_{\infty}.\end{align*}
It follows that if $\gamma<\frac{1}{2\max_{m\in\mathcal{M},n\in\mathcal{N}}\{\theta_{m}B_{m}^{n}\}\max_{n\in\mathcal{N}}\{|\mathcal{N}_{n}|\}}$,
the mapping $(Q_{m}^{n},\forall m\in\mathcal{M},n\in\mathcal{N})$
forms a maximum-norm contraction. \qed

\subsection{Proof of Theorem \ref{thm:For-the-distributed}}\label{proof9}
We first consider the following optimization problem for each user
$n\in\mathcal{N}:$\begin{align}
\max_{\boldsymbol{\sigma}_{n}} & \sum_{m=1}^{M}\sigma_{m}^{n}Q_{m}^{n}(\boldsymbol{P}^{*})-\frac{1}{\gamma}\sum_{m=1}^{M}\sigma_{m}^{n}\ln\sigma_{m}^{n}\label{eq:Ob1}\\
\mbox{subject to} & \sum_{m=1}^{M}\sigma_{m}^{n}=1,\sigma_{m}^{n}\geq0,\forall m\in\mathcal{M}.\nonumber \end{align}
Recall that $Q_{m}^{n}(\boldsymbol{P}^{*})=E[U_{n}(\boldsymbol{a}(T))|\boldsymbol{P}^{*},a_{n}(T)=m]$
denotes the throughput of user $n$ choosing channel $m$ given that other
users adhere to the perceptions $\boldsymbol{P}^{*}$(i.e., other
users\textquoteright{} mixed strategies since $\sigma_{m}^{i*}=\frac{e^{\gamma P_{m}^{i*}}}{\sum_{m'=1}^{M}e^{\gamma P_{m'}^{i*}}},\forall i\neq n$).
Then the objective function (\ref{eq:Ob1}) is to choose a mixed strategy
$\boldsymbol{\sigma}_{n}$ for user $n$ to maximize its expected
throughput $U_{n}(\boldsymbol{\sigma}_{n},\boldsymbol{\sigma}_{-n}^{*})\triangleq\sum_{m=1}^{M}\sigma_{m}^{n}Q_{m}^{n}(\boldsymbol{P}^{*})$
off the term $\frac{1}{\gamma}\sum_{m=1}^{M}\sigma_{m}^{n}\ln\sigma_{m}^{n}$.
The constraint is to ensure that the mixed strategy is feasible.

By the KKT condition, we obtain the optimal solution to (\ref{eq:Ob1})
as $\hat{\sigma}_{m}^{n}=\frac{e^{\gamma Q_{m}^{n}(\boldsymbol{P}^{*})}}{\sum_{m'=1}^{M}e^{\gamma Q_{m'}^{n}(\boldsymbol{P}^{*})}},\forall m\in\mathcal{M}.$
According to (\ref{eq:C2}), we have $Q_{m}^{n}(\boldsymbol{P}^{*})=P_{m}^{n*}$
and hence $\hat{\sigma}_{m}^{n}=\sigma_{m}^{i*}$, which implies that\begin{align*}
& U(\boldsymbol{\sigma}_{n}^{*},\boldsymbol{\sigma}_{-n}^{*})\triangleq\sum_{m=1}^{M}\sigma_{m}^{n*}Q_{m}^{n}(\boldsymbol{P}^{*})
\\ = & \max_{\boldsymbol{\sigma}_{n}}\left\{\sum_{m=1}^{M}\sigma_{m}^{n}Q_{m}^{n}(\boldsymbol{P}^{*})-\frac{1}{\gamma}\sum_{m=1}^{M}\sigma_{m}^{n}\ln\sigma_{m}^{n}\right\}
 +\frac{1}{\gamma}\sum_{m=1}^{M}\sigma_{m}^{n*}\ln\sigma_{m}^{n*}.\end{align*}
It is known from \cite{key-30} that \begin{align*} & \frac{1}{\gamma}\ln\left(\sum_{m=1}^{M}e^{\gamma Q_{m}^{n}(\boldsymbol{P}^{*})}\right) \\
= & \max_{\boldsymbol{\sigma}_{n}}\left\{ \sum_{m=1}^{M}\sigma_{m}^{n}Q_{m}^{n}(\boldsymbol{P}^{*}) -\frac{1}{\gamma}\sum_{m=1}^{M}\sigma_{m}^{n}\ln\sigma_{m}^{n}\right\}. \end{align*}
Since \begin{align*}
  & \max_{\boldsymbol{\sigma}_{n}}U_{n}(\boldsymbol{\sigma}_{n},\boldsymbol{\sigma}_{-n}^{*}) =  \max_{\boldsymbol{\sigma}_{n}}\sum_{m=1}^{M}\sigma_{m}^{n}Q_{m}^{n}(\boldsymbol{P}^{*}) \\
 = & \max_{m\in\mathcal{M}}Q_{m}^{n}(\boldsymbol{P}^{*})\leq \frac{1}{\gamma}\ln\left(\sum_{m=1}^{M}e^{\gamma Q_{m}^{n}(\boldsymbol{P}^{*})}\right),\end{align*}
it follows that\begin{align*}
U(\boldsymbol{\sigma}_{n}^{*},\boldsymbol{\sigma}_{-n}^{*})& =\frac{1}{\gamma}\ln\left(\sum_{m=1}^{M}e^{\gamma Q_{m}^{n}(\boldsymbol{P}^{*})}\right)+\frac{1}{\gamma}\sum_{m=1}^{M}\sigma_{m}^{n*}\ln\sigma_{m}^{n*} \\ &\geq\max_{\boldsymbol{\sigma}_{n}}U_{n}(\boldsymbol{\sigma}_{n},\boldsymbol{\sigma}_{-n}^{*})+\frac{1}{\gamma}\sum_{m=1}^{M}\sigma_{m}^{n*}\ln\sigma_{m}^{n*},\end{align*}
which completes the proof. \qed

\subsection{Proof of Theorem \ref{thm:SINR}} 
We first define that $\tilde{U}_{n}(\boldsymbol{a}))=\frac{\eta_{n}d_{n}^{-\alpha}}{\omega_{0}+\omega_{a_{n}}^{n}+\sum_{i\in\mathcal{N}/\{n\}:a_{i}=a_{n}}\eta_{i}d_{in}^{-\alpha}}$.
According to (\ref{eq:U}), we first have that
\begin{eqnarray*}
 &  & U_{n}(\boldsymbol{a})=\theta W\log_{2}\left(1+\tilde{U}_{n}(\boldsymbol{a}))\right).
\end{eqnarray*}
Since $f(z)=\theta W\log_{2}\left(1+z\right)$ is a a monotonically strictly
increasing function, we have that
\begin{eqnarray*}
 &  & \mbox{sgn}\left(U_{n}(a_{n}^{'},a_{-n})-U_{n}(a_{n},a_{-n})\right)\\
 & = & \mbox{sgn}\left(\tilde{U}_{n}(a_{n}^{'},a_{-n})-\tilde{U}_{n}(a_{n},a_{-n})\right).
\end{eqnarray*}

Suppose a user $k$ changes its channel $a_{k}$ to $a_{k}^{'}$ such
that strategy profile changes from $\boldsymbol{a}$ to $\boldsymbol{a}'$
. We have that
\begin{eqnarray*}
 &  & \Phi(\boldsymbol{a}')-\Phi(\boldsymbol{a})\\
 & = & -\sum_{j\ne k}\frac{\eta_{k}\eta_{j}}{d_{kj}^{\alpha}}I_{\{a_{k}^{'}=a_{j}\}}+\sum_{j\ne k}\frac{\eta_{k}\eta_{j}}{d_{kj}^{\alpha}}I_{\{a_{k}=a_{j}\}}\\
 &  & -\sum_{i\ne k}\frac{\eta_{i}\eta_{k}}{d_{ik}^{\alpha}}I_{\{a_{i}=a_{k}^{'}\}}+\sum_{i\ne k}\frac{\eta_{i}\eta_{k}}{d_{ik}^{\alpha}}I_{\{a_{i}=a_{k}\}}\\
 &  & -2P_{k}\omega_{a_{k}^{'}}^{k}+2P_{k}\omega_{a_{k}}^{k}.
\end{eqnarray*}
Since $d_{ij}$ denotes the distance between user $i$ and user $j$,
we have $d_{ij}=d_{ji}$. Thus,
\begin{eqnarray*}
 &  & \Phi(\boldsymbol{a}')-\Phi(\boldsymbol{a})\\
 & = & -2\sum_{i\ne k}\frac{\eta_{i}\eta_{k}}{d_{ik}^{\alpha}}I_{\{a_{i}=a_{k}^{'}\}}+2\sum_{i\ne k}\frac{\eta_{i}\eta_{k}}{d_{ik}^{\alpha}}I_{\{a_{i}=a_{k}\}}\\
 &  & -2\eta_{k}\omega_{a_{k}^{'}}^{k}+2\eta_{k}\omega_{a_{k}}^{k}\\
 & = & -2\eta_{k}\left(\sum_{i\ne k:I_{\{a_{i}=a_{k}^{'}\}}}\frac{\eta_{i}}{d_{ik}^{\alpha}}+\omega_{k,a_{k}^{'}}\right)\\
 &  & +2\eta_{k}\left(\sum_{i\ne k:I_{\{a_{i}=a_{k}\}}}\frac{\eta_{i}}{d_{ik}^{\alpha}}+\omega_{k,a_{k}}\right)\\
 & = & 2d_{k}^{\alpha}\left(\sum_{i\ne k:I_{\{a_{i}=a_{k}^{'}\}}}\frac{\eta_{i}}{d_{ik}^{\alpha}}+\omega_{k,a_{k}^{'}}+\omega_{0}\right)\\
 &  & \times\left(\sum_{i\ne k:I_{\{a_{i}=a_{k}\}}}\frac{\eta_{i}}{d_{ik}^{\alpha}}+\omega_{k,a_{k}}+\omega_{0}\right)\\
 &  & \times\left(\frac{\eta_{k}d_{k}^{-\alpha}}{\sum_{i\ne k:I_{\{a_{i}=a_{k}^{'}\}}}\frac{\eta_{i}}{d_{ik}^{\alpha}}+\omega_{a_{k}^{'}}^{k}+\omega_{0}}\right.\\
 &  & -\left.\frac{\eta_{k}d_{k}^{-\alpha}}{\sum_{i\ne k:I_{\{a_{i}=a_{k}\}}}\frac{\eta_{i}}{d_{ik}^{\alpha}}+\omega_{a_{k}}^{k}+\omega_{0}}\right)\\
 & = & 2d_{k}^{\alpha}\left(\sum_{i\ne k:I_{\{a_{i}=a_{k}^{'}\}}}\frac{\eta_{i}}{d_{ik}^{\alpha}}+\omega_{a_{k}^{'}}^{k}+\omega_{0}\right)\\
 &  & \times\left(\sum_{i\ne k:I_{\{a_{i}=a_{k}\}}}\frac{\eta_{i}}{d_{ik}^{\alpha}}+\omega_{a_{k}}^{k}+\omega_{0}\right)\\
 &  & \times\left(\tilde{U}_{k}(a_{k},a_{-k})-\tilde{U}_{k}(a_{k}^{'},a_{-k})\right).
\end{eqnarray*}
It follows that
\begin{eqnarray*}
 &  & \mbox{sgn}\left(\Phi(a_{n}^{'},a_{-n})-\Phi(a_{n},a_{-n})\right)\\
 & = & \mbox{sgn}\left(\tilde{U}_{n}(a_{n}^{'},a_{-n})-\tilde{U}_{n}(a_{n},a_{-n})\right)\\
 & = & \mbox{sgn}\left(U_{n}(a_{n}^{'},a_{-n})-U_{n}(a_{n},a_{-n})\right).
\end{eqnarray*} \qed

\end{document}